\documentclass[aps,prb,twocolumn,superscriptaddress,nobalancelastpage,longbibliography]{revtex4-2}
\usepackage{amsmath}
\usepackage{amsfonts}
\usepackage{amssymb}
\usepackage{graphicx}
\usepackage{xcolor}
\usepackage[colorlinks=True,citecolor=myOrange,linkcolor=myRed,urlcolor=myOrange]{hyperref}
\usepackage{hypcap}
\usepackage{braket}
\usepackage{bm}
\usepackage{bbm}
\usepackage[export]{adjustbox}
\usepackage{verbatim}

\newcommand{\ttau}{\tilde{\tau}}
\newcommand{\ua}{\scriptstyle{\uparrow}}
\newcommand{\da}{\scriptstyle{\downarrow}}

\definecolor{myBlue}{RGB}{31,119,180}
\definecolor{myOrange}{RGB}{255,127,14}
\definecolor{myGreen}{RGB}{44,160,44}
\definecolor{myRed}{RGB}{214,39,40}
\definecolor{myPurple}{RGB}{148,103,189}

\definecolor{otherBlue}{cmyk}{1, 0.53, 0, 0.1}
\definecolor{otherOrange}{cmyk}{0, 0.72, 1.0, 0.06}

\definecolor{anotherBlue1}{rgb}{0.1271049596, 0.4401845444, 0.7074971164}
\definecolor{anotherBlue2}{rgb}{0.0313725490, 0.2897347174, 0.570319108}
\definecolor{anotherBlue3}{rgb}{0.2909803921, 0.5945098039, 0.7890196078}
\definecolor{anotherBlue4}{rgb}{0.4980392156, 0.7254133025, 0.8561322568}
\definecolor{anotherGrey1}{rgb}{0.1790080738, 0.1790080738, 0.1790080738}
\definecolor{anotherGrey2}{rgb}{0.0091041906, 0.0091041906, 0.0091041906}
\definecolor{anotherGrey3}{rgb}{0.3713033448, 0.3713033448, 0.3713033448}
\definecolor{anotherGrey4}{rgb}{0.5344098423, 0.5344098423, 0.5344098423}
\definecolor{anotherOrange1}{rgb}{0.8462745098, 0.28069204152, 0.004106113033}

\makeatletter
\def\p@figure{\color{myBlue}}
\def\p@equation{\color{myRed}}

\makeatother

\begin{document}

\title{Local resonances and parametric level dynamics in the many-body localised phase}

\author{S. J. Garratt}
\affiliation{Theoretical Physics, University of Oxford,\\ Parks Road, Oxford OX1 3PU, United Kingdom}
\author{Sthitadhi Roy}
\affiliation{Theoretical Physics, University of Oxford,\\ Parks Road, Oxford OX1 3PU, United Kingdom}
\affiliation{Physical and Theoretical Chemistry, University of Oxford,
South Parks Road, Oxford OX1 3QZ, United Kingdom}
\author{J. T. Chalker}
\affiliation{Theoretical Physics, University of Oxford,\\ Parks Road, Oxford OX1 3PU, United Kingdom}

\date{\today}
             
\begin{abstract}

By varying the disorder realisation in the many-body localised (MBL) phase, we investigate the influence of resonances on spectral properties. The standard theory of the MBL phase is based on the existence of local integrals of motion (LIOM), and eigenstates of the time evolution operator can be described as LIOM configurations. We show that smooth variations of the disorder give rise to avoided level crossings, and we identify these with resonances between LIOM configurations. Through this parametric approach, we develop a theory for resonances in terms of standard properties of non-resonant LIOM. This framework describes resonances that are locally pairwise, and is appropriate in arbitrarily large systems deep within the MBL phase. We show that resonances are associated with large level curvatures on paths through the ensemble of disorder realisations, and we determine the curvature distribution. By considering the level repulsion associated with resonances we calculate the two-point correlator of the level density. We also find the distributions of matrix elements of local observables and discuss implications for low-frequency dynamics.
\end{abstract}

\maketitle

\section{Introduction}\label{sec:introduction}
The spectral properties of quantum many-body systems exhibit a remarkable degree of universality. For example, the emergence of thermodynamic equilibrium in isolated systems places strong constraints on the structure of eigenstates; this is the content of the eigenstate thermalisation hypothesis (ETH) \cite{deutsch1991quantum,srednicki1994chaos,rigol2008thermalization,dallesio2016from,deutsch2018eigenstate}. Additionally, systems that thermalise have spectral statistics that match the predictions of random matrix theory (RMT) \cite{mehta2004random,haake1991quantum,stockmann1999chaos} on fine energy scales. Although thermalisation is generic, with sufficiently strong disorder there is an alternative. In the many-body localised (MBL) phase \cite{gornyi2005interacting,basko2006metal,oganesyan2007localization} local observables fail to thermalise, and the eigenstates exhibit striking departures from the ETH \cite{nandkishore2015manybody,abanin2019colloquium}. Additionally, spectra in the MBL phase resemble Poisson processes: there is no repulsion between typical level pairs in the thermodynamic limit.

In disordered systems, one approach for exploring the connection between the structure of eigenstates and the spectral statistics is to vary the disorder realisation. In this way one generates a fictitious dynamics of the spectrum along paths through the disorder ensemble, parametrised by a fictitious `time'. This idea first arose in the context of RMT \cite{dyson1962brownian}; by allowing matrix elements to evolve stochastically, Dyson developed a theory for the dynamics and equilibrium properties of the eigenvalue gas. An alternative is to consider smooth variations of parameters \cite{pechukas1983distribution,yukawa1985new,nakamura1986complete}. In this way one can characterise the avoided crossings that arise during the fictitious dynamics and relate them to physical properties of the system \cite{wilkinson1987narrowly,wilkinson1989statistics,gaspard1990parametric,goldberg1991parametric,zakrzewski1991distributions,zakrzewski1993parametric1,zakrzewski1993parametric2,vonoppen1994exact,fyodorov1995universality}. Approaches based on parametric level dynamics have also been applied to the study of single-particle disordered conductors \cite{szafer1993universal,simons1993universal,simons1993universalities,chalker1996random}, and in the context of the many-body localisation transition \cite{serbyn2016spectral,filippone2016drude,monthus2016level,monthus2017manybody,maksymov2019energy}.


Theories of the MBL phase are generally based on the existence of local integrals of motion (LIOM) \cite{serbyn2013local,huse2014phenomenology,chandran2015constructing,ros2015integrals}. These are an extensive number of quasi-local operators that commute with time evolution. In particular, in the standard setting of disordered spin-1/2 chains, it is generally expected that one can construct one LIOM per site, and that the support of the LIOM decays exponentially with distance from that site. It has become clear, however, that certain physical quantities are controlled by rare resonances \cite{gopalakrishnan2015low,colmenarez2019statistics,villalonga2020eigenstates,crowley2021constructive}. Resonances alter the structure of LIOM, and here the description of the MBL phase must be refined \cite{imbrie2016diagonalization}. 

In this work, we consider the fictitious dynamics of the spectral properties of a MBL system as the disorder realisation is varied. Avoided crossings of the levels arise naturally, and correspond to resonances between LIOM configurations. While for a sufficiently large system resonances are present in almost all disorder realisations that support the MBL phase, our use of a parametric approach provides a convenient way of identifying their distinctive properties against a background of non-resonant LIOM. In this way, we develop a theory for the resonances that is based on properties of that background. Using this theory, and focusing first on the spectral statistics, we calculate the two-point correlator of the level density and the distribution of level curvatures that arises from variations in the disorder realisation. We then calculate distributions of matrix elements of local observables, and the corresponding spectral functions. Our analytic arguments are supported by numerical calculations in a Floquet model for the MBL phase. Prior to a detailed discussion, we first outline our theory and the results.

\begin{figure}
\includegraphics[width=0.4\textwidth]{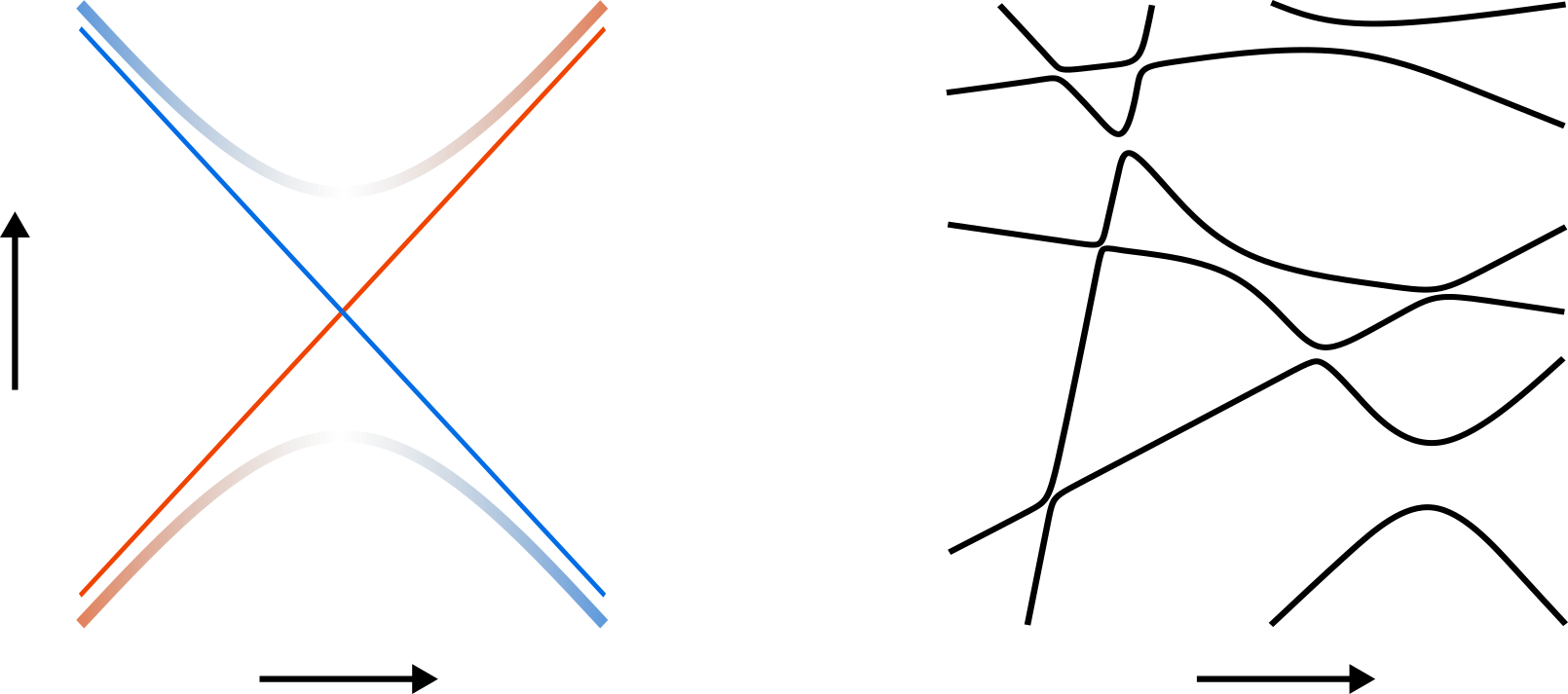}
\put(-220,85){$\theta_2(\lambda_1)$}
\put(-124,85){$\theta_2(\lambda_2)$}
\put(-220,10){$\theta_1(\lambda_1)$}
\put(-124,10){$\theta_1(\lambda_2)$}
\put(2,5){$\theta_1$}
\put(2,40){$\theta_2$}
\put(2,48){$\theta_3$}
\put(2,60){$\theta_4$}
\put(2,70){$\theta_5$}
\put(2,88){$\theta_6$}
\put(-215,47){$\theta$}
\put(-163,-10){$\lambda$}
\put(-40,-10){$\lambda$}
\put(-184,81){\color{otherBlue}$\uparrow\uparrow\uparrow$}
\put(-151,81){\color{otherOrange}$\downarrow\uparrow\downarrow$}
\put(-184,12){\color{otherOrange}$\downarrow\uparrow\downarrow$}
\put(-151,12){\color{otherBlue}$\uparrow\uparrow\uparrow$}
\caption{Cartoons of avoided crossings; $\lambda$ is the fictitious time, and $\theta$ the quasienergy. Left: a pair of levels in the decoupled system (thin lines), where exact degeneracies arise as the disorder realisation is varied, and in the interacting system (thick lines), where they do not. The blue and orange colours represent distinct LIOM configurations, illustrated with coloured vertical arrows. Where the colour fades, the LIOM configuration defined on either side of the crossing are in resonance. Right: a number of levels in the interacting system, highlighting the broad distribution of widths of avoided crossings, as well as the large curvatures near narrow ones.}
\label{fig:crossing}
\end{figure}

\section{Overview \label{sec:overview}}

We study unitary Floquet evolution in spin-1/2 chains. The models we consider offer a simplification relative to Hamiltonian ones in that they have an average level density that is uniform. Additionally, they do not have any conserved densities. The MBL phases that arise in these two classes of models nevertheless share many features \cite{ponte2015manybody,lazarides2015fate,zhang2016floquet}. For integer time $t$ we write our evolution operators as $W^t$ where $W$ is the Floquet operator. Our focus is on the spectral decomposition of $W$, defined by $W\ket{n}=e^{i \theta_n}\ket{n}$. We adopt the convention that the quasienergies $\theta_n$, with $n=1 \ldots 2^L$, are ordered around the unit circle, and that $-\pi < \theta_n \leq \pi$.

\subsection{Floquet model}\label{sec:model}
Throughout this paper, we support our analytic arguments with numerical results based on exact diagonalisation (ED) of Floquet operators. For concreteness we give details on our model here, but we expect our results to apply to MBL systems more generally. Our evolution operator has the structure of a brickwork quantum circuit, so that $W = [\bigotimes_{j \text{ odd}} w_{j,j+1}][\bigotimes_{j \text{ even}} w_{j,j+1}]$ where $j=1 \ldots L$ and
\begin{align}
	w_{j,j+1} = \begin{cases} \exp\big[i\pi J\Sigma_{j,j+1}\big]u_{j}\otimes u_{j+1}, \quad j \, \text{even} \\ 
	\exp\big[i\pi J\Sigma_{j,j+1}\big]u'_{j}\otimes u'_{j+1}, \quad j \, \text{odd}\end{cases}
\label{eq:twositegate}
\end{align}
Here $u_j$ and $u'_j$ are independent Haar-random $2 \times 2$ unitary matrices that represent the precession of spins in on-site fields, and $\Sigma_{j,j+1}$ is the swap operator acting on sites $j$ and $j+1$. We use periodic boundary conditions, and this necessitates $L$ even. Graphically,
\begin{equation}
	W = \includegraphics[valign=c,width=0.3\textwidth]{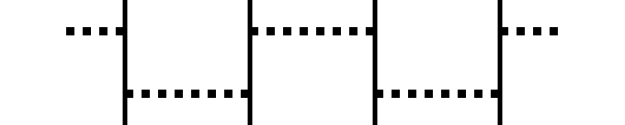}
	\put(-80,17){$J$}
	\put(-110,0){$J$}
	\put(-50,0){$J$}
	\put(-90,-12){$u$}
	\put(-90,2){$u'$}
\end{equation}
where time runs vertically and space horizontally. The solid lines represent the evolution of the different sites under the on-site fields, and dashed horizontal lines represent intersite couplings with strength $J$. Due to the random fields, our model does not have time-reversal symmetry (TRS). We are concerned with weak coupling $J \leq J_c$, with $J_c \simeq 0.07$, where the model appears to be MBL for the accessible range of system sizes \cite{garratt2020manybody}.

We study fictitious dynamics through the ensemble of local disorder realisations by varying the Floquet operator, and we denote by $\lambda$ the fictitious time. Then, for example, $W^t(\lambda)$ is the evolution operator for (integer) time $t$ and at fictitious time $\lambda$. The fictitious dynamics is specified by
\begin{align}
	W(\lambda)=e^{i\lambda G}W,
\label{eq:Wlambda}
\end{align}
where the generator
\begin{align}
	G = \sum_j \vec{v}_j \cdot \vec{\sigma}_j,
\label{eq:Gperturb}
\end{align}
and $\vec{v}_j$ are random unit vectors that are uncorrelated with one another and are independent of $\lambda$. We are concerned with random disorder realisations $W$ and smooth rotations of the fields [Eqs.~\eqref{eq:Wlambda} and \eqref{eq:Gperturb}] with $|\lambda| \ll 1$. For the spectral properties, we often use the notation $W(\lambda)\ket{n(\lambda)}=e^{i\theta_n(\lambda)}\ket{n(\lambda)}$. Wherever we omit the argument $\lambda$, we refer to $\lambda=0$.

In the decoupled limit ($J=0$) and for $\lambda=0$ the evolution operator for site $j$ can be written $u'_j u_j = e^{i\vec{h}_j\cdot\vec{\sigma}_j}$ up to an overall phase, where $\vec{\sigma}_j$ is a vector of standard Pauli matrices. It is convenient to define rotated Pauli matrices $\tau^z_j = (\vec{h}_j/h_j)\cdot \vec{\sigma}_j$, where $h_j \equiv |\vec{h}_j|$, as well as $\tau^{x,y}_j$ chosen so that $[\tau^{\alpha}_j,\tau^{\beta}_j]=2i \varepsilon^{\alpha \beta \gamma} \tau^{\gamma}_j$. The eigenstates of the decoupled model are tensor products of $\tau^z_j$ eigenstates: we have $\tau^z_j\ket{n}=s_{j}\ket{n}$ where $s_{j}=\pm 1$ depends on $\ket{n}$. From Eqs.~\eqref{eq:Wlambda} and \eqref{eq:Gperturb} we see that the directions of the fields vary with $\lambda$, so more generally we write $\vec{h}_j(\lambda)$, and define operators $\tau^{\alpha}_j(\lambda)$ with respect to $\vec{h}_j(\lambda)$. Certain properties of the system for small $J<J_c$ resemble the decoupled limit, as we now discuss.

\subsection{Local integrals of motion}
The standard phenomenology of the MBL phase~\cite{serbyn2013local,huse2014phenomenology,chandran2015constructing,ros2015integrals} tells us that in an $L$-site system there exist $L$ LIOM that commute with one another and with $W$. We denote these operators $\ttau^z_j$, and their eigenvalues $\tilde s_j = \pm 1$. For small $J$, $\ttau^z_j$ is closely related to $\tau^z_j$. More specifically, the LIOM $\ttau^z_j$ can be expressed as a sum of strings of $\tau^{\alpha}_k$ operators, with exponentially decaying support in space away from site $j$ \cite{serbyn2013local,huse2014phenomenology}. 

To construct an operator basis involving $\ttau^z_j$ it is useful to define $\ttau^{x,y}_j$ with similar spatial structure to $\ttau^z_j$, and with $[\ttau^{\alpha}_j,\ttau^{\beta}_k] = 2i\delta_{jk}\epsilon^{\alpha \beta \gamma} \ttau^{\gamma}_j$. Inverting the expansion of $\ttau^{\alpha}_j$ (in terms of $\tau^{\beta}_k$ operators) we anticipate
\begin{align}
	\tau^{\alpha}_j = \sum_{n=1}^{L}\sum_{j_1 \ldots j_n} \sum_{\alpha_1 \ldots \alpha_n} a_{j;j_1 \ldots j_n}^{\alpha;\alpha_1 \ldots \alpha_n} \ttau^{\alpha_1}_{j_1} \ldots \ttau^{\alpha_n}_{j_n}.
\label{eq:LIOMexpansion}
\end{align}
The coefficients $a_{j;j_1 \ldots j_n}^{\alpha;\alpha_1 \ldots \alpha_n}$ describe $n$-body terms with $j_1 < \ldots < j_n$, and it is generally expected that $a_{j;j_1 \ldots j_n}^{\alpha;\alpha_1 \ldots \alpha_n} \sim e^{-|j_1-j_n|/\zeta}$, where $\zeta$ is a decay length that is zero at $J=0$ and that increases with $J$. Hence, for $J \to 0$ we have $\tilde \tau^{\alpha}_j \to \tau^{\alpha}_j$. Note that in reality there is a distribution of $\zeta$, although we neglect this aspect of the problem.

Because the Floquet operator commutes with $\ttau^z_j$ and describes local interactions of the $\tau^{\alpha}_j$ degrees of freedom, it takes the form
\begin{align}
	W = \exp i\Big[ \sum_{n=1}^L \sum_{j_1 \ldots j_n} c_{j_1 \ldots j_n} \ttau^z_{j_1} \ldots \ttau^z_{j_n}\Big],
\label{eq:WLIOM}
\end{align}
up to an overall phase. Here $c_{j_1 \ldots j_n}$ is associated with $n$-body interactions between LIOM $j_1 < \ldots < j_n$, and $c_{j_1 \ldots j_n} \sim e^{-|j_1-j_n|/\zeta}$. In the interest of simplicity we have assumed the same decay length $\zeta$ as in Eq.~\eqref{eq:LIOMexpansion}. As $J$ approaches zero, the one-body terms $c_{j_1}$ approach the physical local fields $h_{j_1}$, and many-body terms such as $c_{j_1 j_2}$ approach zero.

By construction the eigenstates of $W$ in Eq.~\eqref{eq:WLIOM} are eigenstates of $\ttau^z_j$. For $J=0$ it is useful to label eigenstates of $W$ by eigenvalues of $\tau^z_j$, and for ${J \neq 0}$ it is useful to label them by eigenvalues $\tilde s_j = \pm 1$ of $\ttau^z_j$. To describe the entire LIOM configuration we use the notation $\bm{\tilde s}$, reserving the notation $\bm{s}$ for $J=0$. For brevity, here we have restricted ourselves to $\lambda=0$, but we expect a similar construction for general $\lambda$.

\subsection{Local resonances}

On varying $\lambda$ at $J \neq 0$ one finds avoided crossings of quasienergies $\theta_n(\lambda)$. Comparing the effect of varying $\lambda$ for $J=0$ and for $J\neq 0$, in Sec.~\ref{sec:fictitious} we show that these avoided crossings correspond to local resonances between LIOM configurations \cite{villalonga2020eigenstates}. This correspondence is illustrated on the left in Fig.~\ref{fig:crossing}. Although the resonances are local in space, they may involve several nearby LIOM (as defined off-resonance). The implication is that at resonances the LIOM are delocalised over several sites \cite{gopalakrishnan2015low}. On the right in Fig.~\ref{fig:crossing} we indicate the broad distribution of crossing widths. Note that the curvature of the levels as a function of fictitious time is maximal near the middle of an avoided crossing. 

From the standard theory of the MBL phase, we expect that if $\ket{n}$ and $\ket{m}$ are described by LIOM configurations that differ only within a region of length $r > 1$ sites, then $\braket{n|G|m} \sim \Omega(r)$ with
\begin{align}
	\Omega(r) \equiv e^{-(r-1)/\zeta}.
\label{eq:bigOmega}
\end{align}
Note that the $r=1$ case corresponds to high-energy single-site physics, and that the associated matrix elements are non-zero even for $J=0$. Resonances on lengthscale $r>1$, which we will refer to as `$r$-resonances', occur between pairs of levels separated in the spectrum by $\omega \sim \Omega(r)$. Moreover, in a spin-1/2 chain the number of states $\ket{m}$ whose LIOM configuration allows for a resonance with $\ket{n}$ over lengthscale $r$ increases in proportion to $\sim 2^r$ for each unit length of the system. 

The above indicates that, within each eigenstate $\ket{n}$, the spatial density of $r$-resonances (at a given value of $\lambda$) is $\rho(r) \sim e^{-(1/\zeta-\ln 2)r}$. Provided the decay length $\zeta < \zeta_c$, where
\begin{align}
	\zeta_c \equiv \frac{1}{\ln 2},
\end{align}
summing $\rho(r)$ over $r$ one finds that the total density is finite. Our theory can only be appropriate for $\zeta < \zeta_c$, and the condition $\zeta = \zeta_c$ is generally expected to define an upper limit on the boundary of the MBL phase although the true boundary is argued to be at a smaller value of $\zeta$~\cite{deroeck2017stability,morningstar2021avalanches}. While the total density of resonances is finite within the MBL phase, the total number of resonances that each eigenstate participates in is nevertheless linear in $L$. Crucially, the typical spatial separation between resonances is large for small $\zeta$. For example, $r$-resonances are typically separated in space by $\rho^{-1}(r) \gg r$. Consequently, distinct local resonances are independent of one another. Although our focus is on behaviour deep within the MBL phase, we expect that our approach is also appropriate close to the transition provided it is restricted to sufficiently low energies and hence to large $r$. We discuss this further in Sec.~\ref{sec:discussion}. 

\subsection{Results}

Our results are summarised as follows. In Sec.~\ref{sec:fictitious} we show that varying the disorder realisation in the MBL phase gives rise to avoided level crossings. This feature of the dynamics has clear signatures in the distribution of level curvatures $\kappa$. Whereas for $J=0$ there are no large values of $\kappa$, the sharp avoided crossings that necessarily arise for $J \neq 0$ give rise to a heavy power-law tail in the $\kappa$ distribution. Then, through explicit simulation of the fictitious dynamics, we follow pairs of levels through avoided crossings and calculate off-diagonal matrix elements of $\tau^z_j$ operators. These matrix elements are exactly zero for $J=0$, but for $J \neq 0$ we show that they are of order unity at avoided crossings. This is because avoided crossings are resonances between LIOM configurations.

In Sec.~\ref{sec:resonances} we develop our theory for local resonances. Our focus is on the spectral properties of evolution operators that act on finite spatial regions (in a Hamiltonian system one would instead focus on local Hamiltonians). We argue that deep within the MBL phase distinct local resonances do not overlap, and can therefore be treated separately. We describe the fictitious dynamics of the spectra of our local evolution operators, and develop a pairwise description of the avoided crossings that arise. We then explain how these ideas can be applied to the spectrum of an arbitrarily large system.

Armed with this description of the resonances, in Sec.~\ref{sec:spectrum} we discuss spectral statistics. We show how resonances manifest themselves in the two-point correlator of the level density, 
\begin{align}
	p_{\omega}(\omega) = 2^{-2L} \sum_{nm} \big\langle \delta_{2\pi}\big( \omega - \theta_n+\theta_m)\big\rangle,
\label{eq:pomegadef}
\end{align}
where $-\pi < \omega \leq \pi$, and the subscript on the $\delta$-function indicates that its argument is defined modulo $2\pi$ on the interval $(-\pi,\pi]$.  The angular brackets denote a disorder average. The correlator Eq.~\eqref{eq:pomegadef} is normalised so that $\int d\omega p_{\omega}(\omega) = 1$. We calculate analytically the form of the deviations from Poisson statistics, which corresponds to $p_{\omega}(\omega)=[2\pi]^{-1}$. In particular, we show that $[2\pi]^{-1}-p_{\omega}(\omega) \sim |\omega^*/\omega|^{\zeta/\zeta_c}$ for $\omega^* \ll |\omega| \ll 1$, where $\omega^* = L^{\zeta_c/\zeta} e^{-L/\zeta}$. The factor $L^{\zeta_c/\zeta}$ has its origin in the translational entropy associated with the different possible spatial locations of resonances. Exact numerical calculations show excellent support for our theory, which we believe is appropriate for arbitrarily large $L$. We then discuss implications for the behaviour of the spectral form factor at late times in the MBL phase. Following this we return to discuss the distribution of level curvatures
\begin{align}
	p_{\kappa}(\kappa) = 2^{-L} \sum_{n} \big\langle \delta\big( \kappa - \partial_{\lambda}^2 \theta_n\big)\big\rangle,
\label{eq:pkappadef}
\end{align}
and we show that $p_{\kappa}(\kappa) \sim L |\kappa|^{-(2-\zeta/\zeta_c)}$ at large $\kappa$. In this way we relate the statistical properties of avoided crossings to the spatial structure of off-resonant LIOM.

In Sec.~\ref{sec:observables} we apply our theory to the behaviour of local observables. A natural way to characterise their dynamics is through the correlation functions $2^{-L}\text{Tr}[\sigma^{\alpha}_j(t) \sigma^{\beta}_j]$. These are straightforwardly related to correlation functions of the $\tau^{\alpha}_j$ operators. We focus on their behaviour in the frequency domain, and so on the spectral functions. The spectral function whose Fourier transform is the autocorrelation function of $\tau^{\alpha}_j$ can be written
\begin{align}
	S^{\alpha}_j(\omega) = 2^{-L} \sum_{nm} |\braket{n|\tau^{\alpha}_j|m}|^2 \delta_{2\pi}\big( \omega - \theta_n + \theta_m\big).
\label{eq:specfundef}
\end{align}
We show that the quantities $|\braket{n|\tau^{z}_j|m}|^2$ are highly sensitive to resonances, and furthermore that on varying $\lambda$ they exhibit peaks that are approximately Lorentzian. We analytically determine the distributions of matrix elements of local observables, and based on this argue that the spectral functions are dominated by resonances. We find $\langle S^{\alpha}(\omega) \rangle \sim |\omega|^{-\zeta/\zeta_c}$ at small $\omega$, in agreement with previous studies \cite{gopalakrishnan2015low,crowley2021constructive}. Strikingly, this is the same $\omega$-dependence as in the two-point correlator of the level density. The same power of $\omega$ appears in both quantities because, on scale $\omega$, both are controlled by $r$-resonances with $r \sim \zeta \ln |\omega|^{-1}$. 

Our analytic calculations suggest particular dependences of $p_{\omega}(\omega)$, $p_{\kappa}(\kappa)$, and $\langle S^z(\omega) \rangle$, on the lengthscale $\zeta=\zeta(J)$. From our numerical calculations, we can therefore infer values of $\zeta(J)$. In Sec.~\ref{sec:zeta} we show that the values of $\zeta(J)$ extracted from the various different physical quantities agree with one another, as required, and have a dependence on $J$ that is consistent with what is expected from perturbation theory. We summarise our work, and discuss related approaches, in Sec.~\ref{sec:discussion}.

\section{Fictitious dynamics}\label{sec:fictitious}

We are concerned with the fictitious dynamics of the spectrum, at fixed $J$, on smooth paths $\vec{h}_j(\lambda)$ through the ensemble of disorder realisations. First we consider a single site. In that case there are two levels that we can label by $s = \pm 1$, having quasienergies $\pm h(\lambda)$. Note that for Haar-random $u,u'$ [Eq.~\eqref{eq:twositegate}] the probability density for a given $h$ vanishes as $\sim h^2$ for small $h$. As $\lambda$ is varied, the levels $s = \pm 1$ typically undergo wide avoided crossings with gaps of order unity.

For $L>1$ sites with $J=0$, on the other hand, the quasienergy associated with configuration $\bm{s}$ is
\begin{align}
	\theta(\lambda;\bm{s}) = \sum_{j=1}^L h_j(\lambda) s_j.
\label{eq:decoupledquasienergies}
\end{align}
If the state labels $\bm{s}$ and $\bm{s'}$ differ on only one site, $j$, the characteristic level separation $2 h_j(\lambda)$ is of order unity. Such pairs of levels undergo wide avoided crossings, as in the single-site problem. By contrast, for $\bm{s}$ and $\bm{s'}$ differing on multiple sites, for $J=0$ we generically find crossings of $\theta(\lambda;\bm{s})$ and $\theta(\lambda;\bm{s'})$ as $\lambda$ is varied. Note that, without fine-tuning, these crossings are all pairwise: no three $\theta(\lambda;\bm{s})$ meet at a point. Because we label eigenstates $\ket{n}$ by their order in the spectrum, i.e. $\theta_n < \theta_{n+1}$, at these crossings the configurations $\bm{s}$ associated with the different $\theta_n$ are exchanged. This convention is indicated by the thin lines on the left in Fig.~\ref{fig:crossing}, and will prove useful in the following.

For small $J \neq 0$, the fictitious dynamics is altered drastically: with probability one there are no exact degeneracies of the $\theta_n(\lambda)$ as $\lambda$ is varied \footnote{Note that without TRS, three parameters in the evolution operator must simultaneously be tuned to zero in order for two levels to be degenerate}. This is because the interactions give rise to non-zero off-diagonal matrix elements $\braket{n|G|m}$. As a result, gaps open where for $J=0$ there were exact crossings in the $\lambda$-$\theta$ plane, as illustrated by the thick lines on the left in Fig.~\ref{fig:crossing}. At non-zero $J$, and for general smooth variations of the $\vec{h}_j(\lambda)$, the $\theta_n(\lambda)$ follow smooth paths and undergo a series avoided crossings as they do so. 

\begin{figure}
\includegraphics[width=0.47\textwidth]{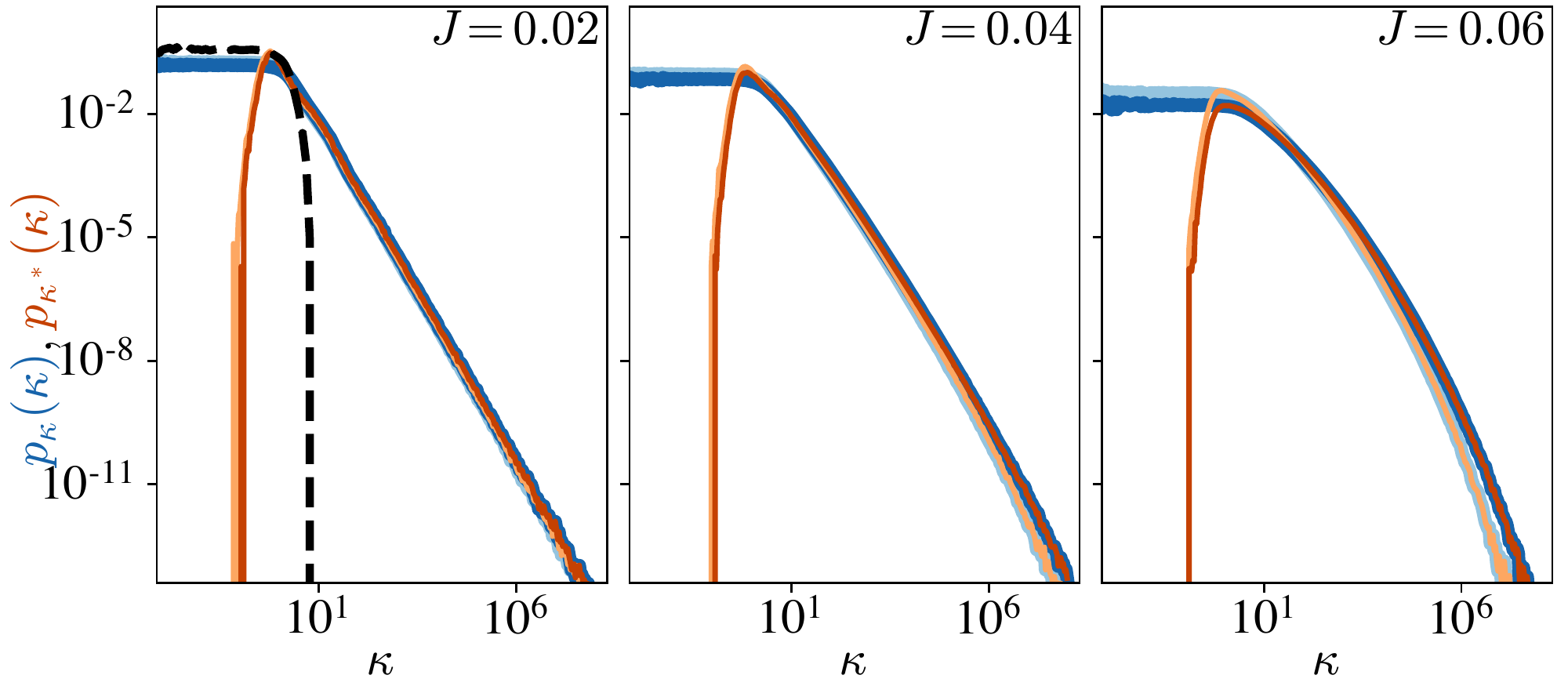}
\caption{Distributions of curvatures $\kappa$ (blues) and $\kappa^*$ (oranges), the latter defined as the largest term in the sum in Eq.~\eqref{eq:curvature}. We show various $J$ (panels), $L=12$ (light) and $L=14$ (dark). The dashed black line shows $p_{\kappa}(\kappa)$ for $J=0$.}
\label{fig:curvature1}
\end{figure}

As an objective way of characterising avoided crossings, independent of any choice for the LIOM description, we consider the curvatures $\kappa_n \equiv \partial_{\lambda}^2 \theta_n$. For our model they are given by
\begin{align}
	\kappa_n = \sum_{m \neq n} |\braket{m|G|n}|^2 \cot[(\theta_n-\theta_m)/2],
\label{eq:curvature}
\end{align}
which follows from perturbation theory for unitary operators as opposed to Hermitian ones [see Appendix~\ref{app:perturb}]. For $J=0$, exact level crossings occur, and there are no large values of $\kappa$. For $J \neq 0$, crossings are avoided, and there we expect large $\kappa$ as illustrated in Fig.~\ref{fig:crossing}.

In Fig.~\ref{fig:curvature1} we show the distribution of curvatures $p_{\kappa}$. For $J=0$ there is no weight at large $\kappa$, whereas for small $J \neq 0$ a heavy tail develops: $p_{\kappa}(\kappa) \sim 1/|\kappa|^{\gamma}$ with $1 < \gamma < 2$, as we explain in Sec.~\ref{sec:curvatures}. Deviations from a power law are evident close to the transition (for example at $J = 0.06$), but this regime is not our focus. We note that heavy tails in the distribution of curvatures have been identified previously in Refs.~\cite{filippone2016drude,monthus2017manybody,maksymov2019energy}. In the ergodic phase one instead expects $p_{\kappa}$ that behaves as in RMT: $p_{\kappa}(\kappa) \sim 1/|\kappa|^{\beta+2}$ so that $\gamma=\beta+2$, where $\beta=2$ is the level-repulsion exponent for systems without TRS~\cite{gaspard1990parametric,zakrzewski1993parametric1,vonoppen1994exact,fyodorov1995universality}.

Additionally, if avoided crossings are pairwise for ${J \neq 0}$, we expect that when $\kappa$ is large the sum on the right-hand side of Eq.~\eqref{eq:curvature} is dominated by its largest (in magnitude) term, which we denote $\kappa^*$. The pairwise character of the crossings has previously been discussed in Ref.~\cite{villalonga2020eigenstates}, although in Sec.~\ref{sec:resonances} we will argue that the avoided crossings are only locally pairwise, in a sense that we will make clear. It is nevertheless the case that for the range of system sizes that is accessible numerically, the distribution of $\kappa$ matches that of $\kappa^*$ for large curvatures.

\begin{figure}
\includegraphics[width=0.47\textwidth]{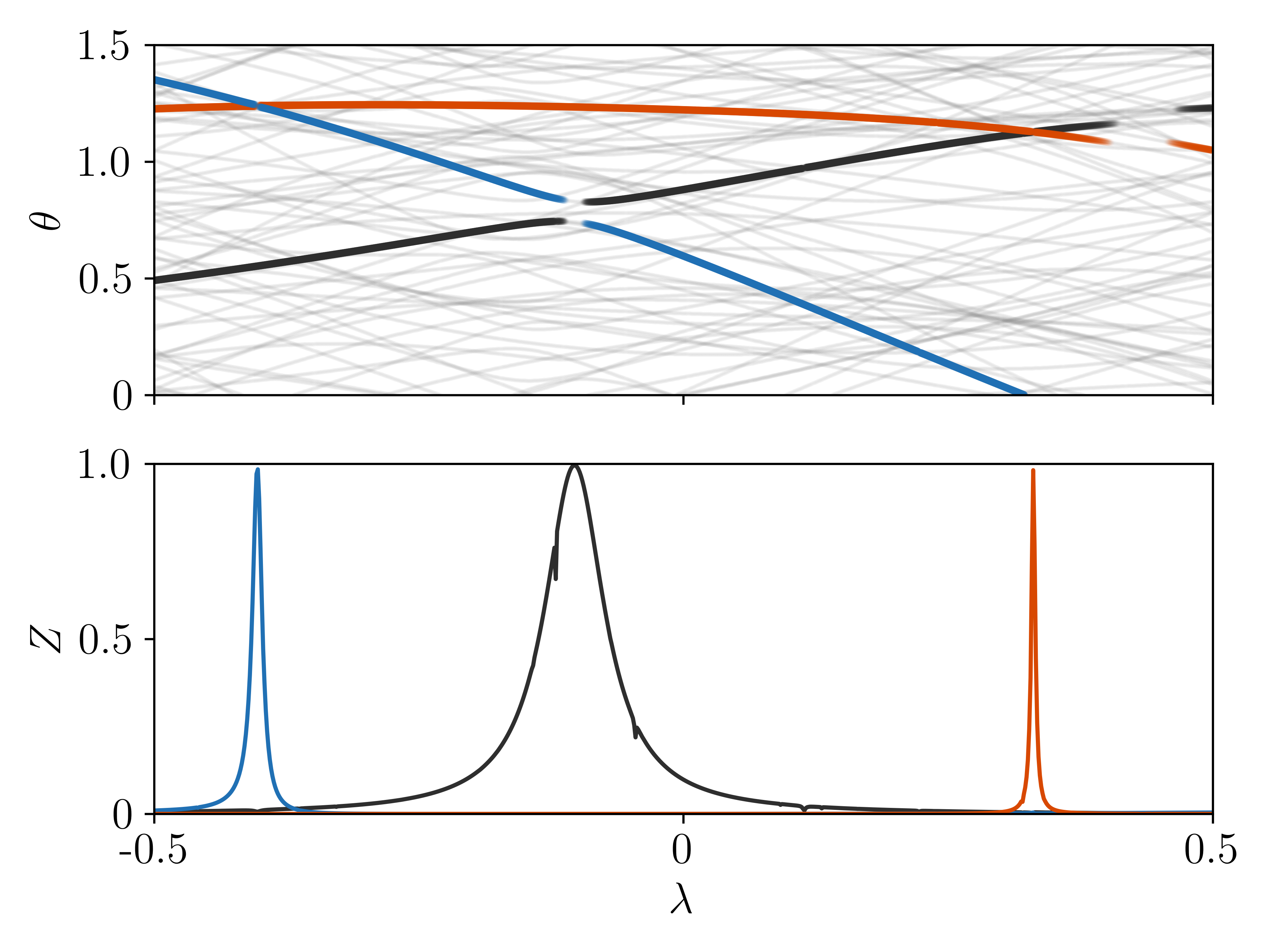}
	\put(-190,124){\color{anotherGrey1}$\ua\ua\ua\da\da\ua\ua\ua$}	
	\put(-184,143){\color{anotherBlue1}$\ua\ua\ua\ua\ua\ua\ua\ua$}
	\put(-135,163){\color{anotherOrange1}$\ua\ua\da\da\ua\ua\ua\ua$}
\caption{Numerical demonstration of fictitious dynamics with $L=8$ and $J=0.02$. Upper: quasienergies $\theta_n(\lambda)$. All levels are shown grey, with three highlighted according to the scheme described in the text. The corresponding LIOM configurations are shown as coloured arrows. Lower: $Z_{nm,j}(\lambda)$ [Eq.~\eqref{eq:bigZeigenstates}] for pairs of levels $\ket{n}$ and $\ket{m}$ highlighted in the upper panel. The choice of $j$ is described in the text. The three peaks are centred on the highlighted crossings, and correspond to $r$-resonances with $r=2,2,3$ in order of increasing $\lambda$.}
\label{fig:leveldynamics}
\end{figure}

Another way to characterise the fictitious dynamics is through the level velocities $\partial_{\lambda}\theta_n$ \cite{maksymov2019energy}. These are given at first order in perturbation theory by $\partial_{\lambda}\theta_n = \braket{n|G|n}$. However, signatures of avoided crossings in the distribution of level velocities are less striking than in that of curvatures. In particular, at $J=0$ one expects Gaussian distributed $\partial_{\lambda}\theta_n$ from the central limit theorem, and this is not substantially altered for small $J \neq 0$.

Before a more detailed discussion of the connection between avoided crossings and resonances, it is helpful to consider a two-level system fine-tuned to have a narrowly-avoided crossing. This system has a different character to the single-site problem discussed at the beginning of this section, where avoided crossings are typically wide. Suppose that here the Floquet operator $W(\lambda) = e^{i (\lambda-\lambda_0) \sigma^z}$, and write $\sigma^z \ket{q} = q \ket{q}$ with $q=\pm 1$. The quasienergies are ${\theta(\lambda;q)= q(\lambda-\lambda_0)}$. As $\lambda$ is varied through $\lambda_0$ there is an exact crossing, and the $\sigma^z$ eigenvalue $q$ `passes through' this crossing. Consider now the case with an additional transverse field. In an abuse of notation, we denote its strength by $\frac{1}{2}\Omega$. The Floquet operator then takes the form $W(\lambda) = e^{i (\lambda-\lambda_0) \sigma^z + i \Omega \sigma^x/2}$, and in the limit $|\lambda-\lambda_0| \gg \Omega > 0$ the eigenstates are again $\ket{q}$ with $q=\pm 1$. However, for $\lambda=\lambda_0$ they are equal amplitude superpositions of the two states $\ket{q}$, and the quasienergy gap is $\omega=\Omega$. We therefore have an avoided crossing, and for $|\lambda-\lambda_0| \lesssim \Omega$ the eigenstates as defined at large $|\lambda-\lambda_0|$ are resonant. Comparing large negative and large positive $(\lambda-\lambda_0)$, the $\sigma^z$ eigenvalue passes through the (avoided) crossing, as with $\Omega=0$.

For our spin chain, and for $J=0$, we have exact crossings, whereas for $J \neq 0$ we have avoided crossings. As $\lambda$ is varied the LIOM configurations pass through the crossings. For $J \neq 0$ and in the vicinity of the avoided crossing, the true eigenstates (locally) resemble superpositions of the LIOM configurations defined away from the crossing: they are resonant. This is illustrated on the left in Fig.~\ref{fig:crossing}, and we discuss this point in more detail in Sec.~\ref{sec:resonances}.

We confirm this picture in the first instance by numerically following a trajectory of the fields $h_j(\lambda)$ for a finite system, using the protocol in Eqs.~\eqref{eq:Wlambda} and \eqref{eq:Gperturb}. To identify the eigenstates that closely resemble a selected set of configurations $\bm{s}$ of the decoupled system, we find the $n$ that minimises $\sum_{j=1}^L |\braket{n(\lambda)|\tau^z_j(\lambda)|n(\lambda)}-s_j|^2$ at each $\lambda$. In this way we can hope to trace out the paths of our selected LIOM configurations as $\lambda$ is varied. A representative result is shown in the upper panel of Fig.~\ref{fig:leveldynamics}, where we highlight three LIOM configurations. We have omitted the colouring near avoided crossings.

Sensitive probes of the resonances are provided by off-diagonal matrix elements of $\tau^z_j$. We define
\begin{align}
	Z_{nm,j}(\lambda) \equiv |\braket{n(\lambda)|\tau^z_j(\lambda)|m(\lambda)}|^2,
\label{eq:bigZeigenstates}
\end{align}
and in the lower panel of Fig.~\ref{fig:leveldynamics} we show $Z_{nm,j}(\lambda)$ for the pairs of levels highlighted in the upper panel. There we choose $j$ to be a site at which $s_j \neq s_j'$, where the configurations $\bm{s}$ and $\bm{s'}$ are respectively associated with the eigenstates $\ket{n(\lambda)}$ and $\ket{m(\lambda)}$ via the scheme described in the previous paragraph. Away from avoided crossings the LIOM $\ttau^z_j(\lambda)$ closely resemble the physical operators $\tau^z_j(\lambda)$, so the eigenstates of $W(\lambda)$ have small off-diagonal matrix elements of $\tau^z_j(\lambda)$ operators. It is clear from the lower panel of Fig.~\ref{fig:leveldynamics} that this is not the case at avoided crossings: there, $Z_{nm,j}(\lambda)$ is large. This indicates that avoided crossings are resonances, and we elaborate on this in Sec.~\ref{sec:observables}.

\section{Description of resonances}\label{sec:resonances}

From the connection between resonances and avoided crossings, we now develop a local description of these phenomena that is based on standard properties of LIOM in non-resonant regions. First, in Sec.~\ref{sec:local}, we discuss why a local description of a single resonance is possible. In Sec.~\ref{sec:density} we show that the resonances are rare for small $\zeta$, and determine their density in space. In Sec.~\ref{sec:pairwise} we develop a pairwise description of a local resonance, and in Sec.~\ref{sec:separation} we discuss how to apply our model to multiple resonances in large systems. In Sec.~\ref{sec:averaging} we consider the steps involved in ensemble averaging the properties associated with resonances.

\subsection{Local degrees of freedom}\label{sec:local}

Here we argue that a description for the resonances should start from evolution operators that act on finite subregions. It is necessary to first discuss the various quasienergy scales involved in the problem. We start by identifying a pair of eigenstates of the LIOM, at fictitious time $\lambda=0$, that are not involved in any resonances. We denote these by $\ket{\bm{\tilde s}}$ and $\ket{\bm{\tilde s'}}$. Although we will soon see that this is only possible with small $L$, these states are a useful theoretical tool. As we vary the disorder realisation, we suppose that these states pass through a resonance.

In the first instance, we must ask how small the quasienergy separation between our states must be for us to observe resonant behaviour. This scale is set by the off-diagonal matrix elements of the generator $G$ of the fictitious dynamics, $\braket{\bm{\tilde s}|G|\bm{\tilde s'}}$. From Eqs.~\eqref{eq:Gperturb} and \eqref{eq:LIOMexpansion} we have an expression for $G$ as a sum over strings of $\ttau^{\alpha}_j$ operators, with the weights of the strings decaying exponentially with their spatial extent. If $\ket{\bm{\tilde s}}$ and $\ket{\bm{\tilde s'}}$ have the same LIOM configuration within a region of at most $L-r$ contiguous sites, the dominant contributions to $\braket{\bm{\tilde s}|G|\bm{\tilde s'}}$ come from $\ttau^{\alpha}_j$ strings with length $r$. That is, if there exists $k$ such that $\tilde s_j = \tilde s'_j$ for $k \leq j \leq l$ with $l=k+L-r-1$, but $\tilde s_k \neq \tilde s'_k$ and $\tilde s_l \neq \tilde s'_l$, we expect
\begin{align}
	\braket{\bm{\tilde s}|G|\bm{\tilde s'}} \sim r^{1/2} e^{-(r-1)/\zeta},
\label{eq:offdiagG}
\end{align}
which is on energy scale $\Omega(r) = e^{-(r-1)/\zeta}$. The factor $r^{1/2}$ is implied at large $r$ by the central limit theorem, but for simplicity we neglect these factors from here on. If on varying $\lambda$ our states are brought within a quasienergy separation $|\omega| \lesssim |\braket{\bm{\tilde s}|G|\bm{\tilde s'}}|$ of one another, we expect an $r$-resonance between them. Note that this application of the LIOM picture to estimate the magnitudes of matrix elements of $G$ relies on the fact that $\ket{\bm{\tilde s}}$ and $\ket{\bm{\tilde s'}}$ are eigenstates of $W(\lambda)$ for $\lambda=0$, and so are independent of $G$.

To describe this $r$-resonance we neglect energy scales of order $\Omega(r+1)$ and smaller. Because of this, we need only consider degrees of freedom that reside in an interval of length $b r$ sites centred on the resonance, with $b \sim 3$. This includes a `buffer' region of $\sim r$ sites on each end \cite{imbrie2016diagonalization,deroeck2017manybody}. This buffer region is necessary because, although the resonance of interest is only within the central region of length $r$, the degrees of freedom involved are coupled to those in the buffer on energy scales $\Omega(r)$ and above.

For this reason, in describing the $r$-resonance, we focus on the evolution operator for a finite region of $br$ sites centred on it. Equivalently, in a Hamiltonian model, we can consider the Hamiltonian for this part of the system. Note that if we did not shift our focus to local operators in this way, discussions of resonances in large systems would involve considering sets of coupled states that are exponentially large in system size (suppose that a given state in a large system is involved in $N$ resonances; the states involved in these resonances span a space of dimension $2^{N}$). 

We denote the Floquet operator for the region of $br$ sites by $W_{br}(\lambda)$, with $W_{br} \equiv W_{br}(0)$. Such an operator can be obtained by discarding all terms in Eq.~\eqref{eq:WLIOM} that involve operators outside of our region of $br$ sites. This corresponds to a model for the $r$-resonance that is based on the approximation
\begin{align}
	W(\lambda) \simeq W_{br}(\lambda) \otimes W_{L-br}(\lambda),
\label{eq:approxdecomposition1}
\end{align}
where $W_{L-br}(\lambda)$ acts only on the complement of the region of $br$ sites. Deep in the MBL phase we expect that a decomposition of the form in Eq.~\eqref{eq:approxdecomposition1} is sufficient to describe statistical properties of the resonance on quasienergy scales $|\omega| \gtrsim \Omega(r)$. Because resonances in different spatial locations generally occur over different intervals in $\lambda$, the structure of the tensor product decomposition of $W(\lambda)$ that is required itself depends on $\lambda$. Note also that, for $r > L/b$, it is not meaningful to discuss a buffer region of $br$ sites. In that case the notation $W_{br}(\lambda)$ should be understood to refer to the Floquet operator $W(\lambda)$ for the full system.

\subsection{Finite density of resonances}\label{sec:density}
A local description of resonances is simplified when they can be treated independently. To see how resonances can be rare in space despite the fact that the level density grows exponentially with $L$, note that for a pair of LIOM configurations $\ket{\bm{\tilde s}}$ and $\ket{\bm{\tilde s'}}$ to be resonant the corresponding quasienergies should be within $\Omega(r)$ of each other, and that $\Omega(r)$ decays exponentially with $r$. If the quasienergies are to a first approximation uncorrelated, the probability for $\ket{\bm{\tilde s}}$ and $\ket{\bm{\tilde s'}}$ to be close enough in quasienergy to form a resonance is $\sim [2\pi]^{-1}\Omega(r)$. 

We denote by $p_r(r)$ the probability that a randomly selected configuration $\bm{\tilde s'}$ is the same as $\bm{\tilde{s}}$ over a region of maximum length $L-r$. The number of such $\bm{\tilde{s}'}$ is $2^L p_r(r)$. We have $2^L p_r(0)=1$ and $2^L p_r(1)=L$, while for $1 < r \leq L/2$ we find, with periodic boundary conditions,
\begin{align}
	2^L p_r(r) = L 2^{r-2}.
\label{eq:pr}
\end{align}
For $r > L/2$, $2^L p_r(r)$ is upper bounded by $L 2^{r-2}$, and for large $L$ Eq.~\eqref{eq:pr} remains a useful approximation until ${r \simeq L}$. We then find, for example, ${2^L p_r(L) = 1}$. Note that the distribution $p_r(r)$ is normalised as ${\sum_{r=0}^L p_r(r)=1}$. The average spatial density of $r$-resonances involving a given $\bm{\tilde s}$ is
\begin{align}
	\rho(r) &= [2\pi]^{-1} 2^{r-2} e^{-(r-1)/\zeta},
\label{eq:rhor}
\end{align}
i.e. $\rho(r) \sim e^{-(1/\zeta-1/\zeta_c)r}$. Eq.~\eqref{eq:rhor} is strictly valid only for $1 < r \leq L/2$, and takes a more complicated form for $r > L/2$. From Eq.~\eqref{eq:rhor}, the overall spatial density of many-body resonances is
\begin{align}
	\rho = \sum_{r=2}^L \rho(r),
\label{eq:resonancedensity}
\end{align}
which is finite for all $L$ provided $\zeta < \zeta_c$, and goes to zero as $\zeta \to 0$. This means that a typical eigenstate participates in $\sim \rho L$ resonances. From Eq.~\eqref{eq:rhor} we find that in a given eigenstate the fraction of the chain involved in $r$-resonances is $r\rho(r)$. For all $r$, this is small for small $\zeta$, which means that deep in the MBL phase distinct $r$-resonances do not overlap in space. In fact, for any $\zeta < \zeta_c$, at sufficiently large $r$ we again find small $r \rho(r)$.

\begin{figure}
	\includegraphics[width=0.47\textwidth]{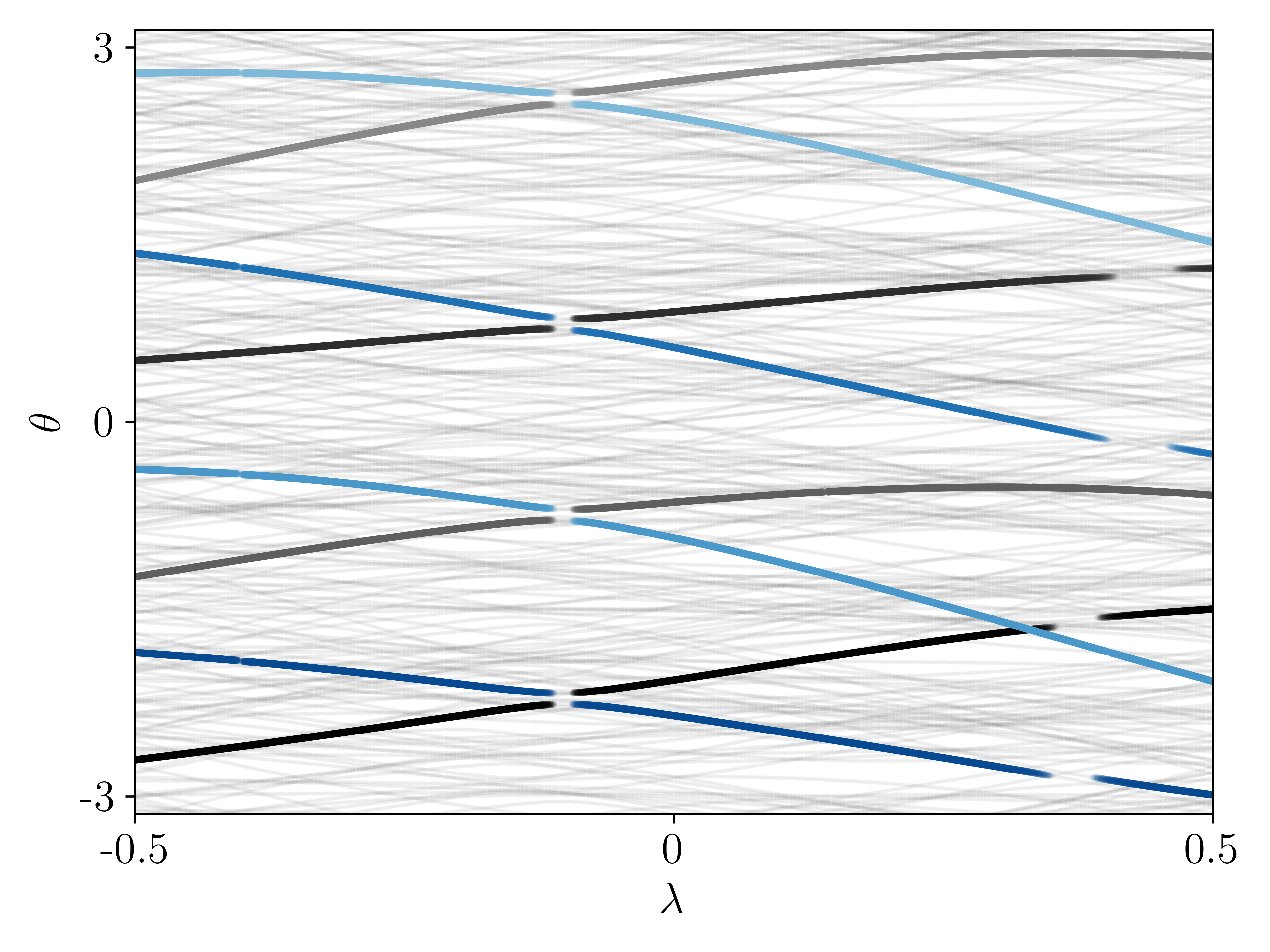}
	\put(-95,128){\color{anotherGrey1}$\ua\ua\ua\displaystyle{\downarrow\downarrow}\ua\ua\ua$}	
	\put(-95,112){\color{anotherBlue1}$\ua\ua\ua\displaystyle{\uparrow\uparrow}\ua\ua\ua$}
	\put(-197,31){\color{anotherGrey2}$\da\ua\ua\displaystyle{\downarrow\downarrow}\ua\ua\ua$}	
	\put(-197,45){\color{anotherBlue2}$\da\ua\ua\displaystyle{\uparrow\uparrow}\ua\ua\ua$}
	\put(-201,80){\color{anotherGrey3}$\ua\ua\ua\displaystyle{\downarrow\downarrow}\ua\ua\da$}	
	\put(-201,93){\color{anotherBlue3}$\ua\ua\ua\displaystyle{\uparrow\uparrow}\ua\ua\da$}
	\put(-210,140){\color{anotherGrey4}$\da\ua\ua\displaystyle{\downarrow\downarrow}\ua\ua\da$}	
	\put(-210,158){\color{anotherBlue4}$\da\ua\ua\displaystyle{\uparrow\uparrow}\ua\ua\da$}
\caption{Fictitious dynamics on the same path through the disorder ensemble as in Fig.~\ref{fig:leveldynamics}, showing copies of an $r=2$ resonance involving the central two sites. LIOM configurations associated with highlighted paths are indicated by vertical arrows. Avoided crossings at $\lambda \approx -0.1$ are between LIOM configurations that differ over the central $r=2$ sites (large arrows). Different shades correspond to LIOM configurations differing far from the central sites.}
\label{fig:copies}
\end{figure}
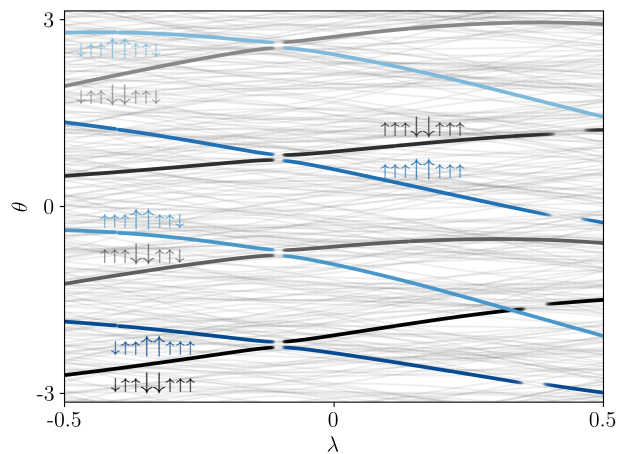

The implication is that, under the conditions just described, resonances can be treated as pairwise, but only locally. Each eigenstate of $W_{br}(\lambda)$ in Eq.~\eqref{eq:approxdecomposition1} is typically involved in no more than one $r$-resonance. But, in a large system, we expect $\sim \rho L$ resonances in each eigenstate of $W(\lambda)$. This situation can be modelled by further decomposing $W_{L-br}(\lambda)$ in Eq.~\eqref{eq:approxdecomposition1}. In this way our approximate model for $r$-resonances in a large system amounts to a tensor decomposition of $W(\lambda)$ into (i) Floquet operators of the form $W_{br}(\lambda)$, that act on resonant regions, and (ii) Floquet operators that act on the non-resonant regions between them. 

The decomposition in Eq.~\eqref{eq:approxdecomposition1} suggests that each resonance in the spectrum of $W_{br}(\lambda)$ appears $2^{L-br}$ times in the spectrum of $W(\lambda)$. It is straightforward to confirm the existence of these `copies' of the resonance using the scheme used to generate Fig.~\ref{fig:leveldynamics}. Following the same path through the ensemble of disorder realisations, in Fig.~\ref{fig:copies} we show four copies of the ($r=2$) resonance that in Fig.~\ref{fig:leveldynamics} is centred on $\lambda \approx -0.1$. If Eq.~\eqref{eq:approxdecomposition1} were exact, the central $\lambda$-coordinates of the different copies of the resonance would be the same. Of course, in reality they are shifted with respect to one another, but these shifts are much smaller than the widths of the resonance in $\lambda$. This is because the resonant LIOM are weakly coupled to degrees of freedom outside of the region of length $br$. We now develop a description of an individual $r$-resonance, and so consider an operator $W_{br}(\lambda)$.

\subsection{Pairwise model for a local resonance}\label{sec:pairwise}
In this section we consider a resonance on a particular lengthscale $r > 1$, so our focus is on the behaviour of $W_{br}(\lambda)$ as $\lambda$ is varied. We have 
\begin{align}
	W_{br}(\lambda)=e^{i \lambda G_{br}}W_{br},
	\notag
\end{align}
where $G_{br}$ is a sum of local Hermitian operators acting only in the resonant region [i.e. it corresponds to a subset of the terms in Eq.~\eqref{eq:Gperturb}]. For simplicity we refer to the quasienergies of $W_{br}(\lambda)$ as $\theta_n(\lambda)$, and where we use the notation $\ket{\bm{\tilde{s}}}$, we refer to the LIOM configuration within the region of length $br$. Note that here the Hilbert space of interest is only of dimension $2^{br}$. 

A description of the resonance can be formulated in terms of the spectrum of $W_{br}(\lambda)$ and the fictitious time evolution operator $U_{br}(\lambda,\lambda')$ for its eigenstates. This is defined by its matrix elements
\begin{align}
	[U_{br}(\lambda,\lambda')]_{mn} = \braket{m(\lambda)|n(\lambda')},
\label{eq:Ulambda}
\end{align}
and we first discuss its behaviour for $J=0$. In that case the eigenstates are product states of the $\tau^z_j(\lambda)$ operators, and $U_{br}(\lambda,\lambda')$ captures changes in the eigenstates that arise from rotations of $\tau^z_j(\lambda)$. If we choose $\lambda'$ and $\lambda$ to lie on opposite sides of an exact level crossing, then in the limit of small $|\lambda-\lambda'|$, $U_{br}(\lambda,\lambda')$ acts as a swap operation on the crossing level pair. Changes in the operators $\tau^z_j(\lambda)$ instead occur over fictitious time intervals that are of order unity. The operator $U_{br}(\lambda,\lambda')$ therefore describes changes in the operators $\tau^z_j$ that are `slow' in fictitious time, and exact crossings of levels that are instantaneous. For small $J \neq 0$ we argue that a similar situation arises because avoided crossings corresponding to $r$-resonances take place over intervals in $\lambda$ that are of order $\Omega(r) \ll 1$. In the following we neglect the `slow' changes in the LIOM $\tilde \tau^z_j(\lambda)$ that occur in the intervals between successive resonances.

For $J=0$, crossings are pairwise with probability one, and the important simplification for small $J \neq 0$ is that a pairwise description of local resonances is possible. This is because it is unlikely for a given LIOM configuration in our finite region to be involved in even one $r$-resonance, as discussed in Sec.~\ref{sec:density}. As a result, $U_{br}(\lambda,\lambda')$ has a similar structure for small $J \neq 0$ as for $J=0$.

To describe the resonance for small $J \neq 0$ we proceed as follows. First, at the reference point $\lambda=0$, we identify a pair of eigenstates of $W_{br}$. With high probability at small $J$, these states will not be involved in an $r$-resonance. We label them by $\ket{\bm{\tilde s}}$ and $\ket{\bm{\tilde s'}}$, and consider the situation where, at $\lambda > 0$, the eigenstates pass through an $r$-resonance. For ease of presentation we suppose that $\ket{\bm{\tilde s}}$ and $\ket{\bm{\tilde s'}}$ are neighbours in the spectrum of $W_{br}$, although later we will see that this is too strong a restriction.

For the resonance of interest we restrict ourselves to an effective description within the two-dimensional space spanned by $\ket{\bm{\tilde s}}$ and $\ket{\bm{\tilde s'}}$. Denoting by $\omega_{br}(\lambda)$ their quasienergy separation, which we here choose to be positive for convenience, we must have $\omega_{br}(0) \gg \Omega(r)$ if our states are not resonant at $\lambda=0$. From Eq.~\eqref{eq:offdiagG} it is natural to expect on varying $\lambda$ the minimum separation of the levels at the resonance is on the scale $\Omega(r)$. Additionally, far from the resonance, the magnitude of the level velocity ${|\partial_{\lambda}\omega(\lambda)| \sim 1}$ up to factors of order $r^{1/2}$, as for $J=0$. This suggests that the resonance is centred on fictitious time $\lambda_0 \sim \omega_{br}(0)$. As we discuss in Appendix~\ref{app:pairwise}, the quasienergy splitting takes the Landau-Zener form
\begin{align}
	\omega_{br}(\lambda;\lambda_0,z,r) = \sqrt{(\lambda-\lambda_0)^2 + |z|^2 \Omega^2(r)}.
\label{eq:splitting}
\end{align}
Here $z$ is a complex number of order unity, and we have made the parametric dependences on $z$, $\lambda_0$, and $r$ explicit.

We can similarly follow the eigenstates in fictitious time. We find
\begin{align}
\ket{+(\lambda)} &= \cos[\varphi(\lambda)/2]\ket{\bm{\tilde s}}+\sin[\varphi(\lambda)/2]\ket{\bm{\tilde s'}} \notag \\
\ket{-(\lambda)} &= \cos[\varphi(\lambda)/2]\ket{\bm{\tilde s'}}
-\sin[\varphi(\lambda)/2]\ket{\bm{\tilde s}},
\label{eq:plusminus}
\end{align}
where for $\omega(0) \gg \Omega(r)$, from Eq.~\eqref{eq:cotphi},
\begin{align}
	\tan \varphi(\lambda;\lambda_0,z,r) = \frac{|z|\Omega(r)}{\lambda_0-\lambda}.
\label{eq:tanphi}
\end{align}
As $\lambda-\lambda_0$ varies from large negative values to large positive ones, $\varphi(\lambda)$ increases from $\varphi(\lambda) \simeq 0$ to $\varphi(\lambda) \simeq \pi$.

The matrix elements of $U_{br}(\lambda,\lambda')$ corresponding to this two-dimensional space can be determined from Eq.~\eqref{eq:plusminus}. For $\lambda$ and $\lambda'$ far from $\lambda_0$, with $\lambda' < \lambda_0 < \lambda$, the matrix $U_{br}(\lambda,\lambda')$ describes a swap of the eigenstates. In this way LIOM configurations are exchanged from one side of the avoided crossing to the other, as discussed in Sec.~\ref{sec:fictitious}. On the other hand, at $\lambda=\lambda_0$ the eigenstates are equal-amplitude superpositions of those at large $\lambda$. This is the middle of the resonance. 

Note that, if there are no other resonances, then at this level of approximation $U_{br}(\lambda,\lambda')$ acts as the identity in the complement of the two-dimensional space discussed above. This complement has dimension $2^{br}-2$.

\subsection{Separation of scales}\label{sec:separation}
In Sec.~\ref{sec:local} we have argued that in order to describe an $r$-resonance, it is sufficient to consider the dynamics of degrees of freedom in a region of $\sim br$ sites. Because a local description is possible, in Sec.~\ref{sec:pairwise} we have considered the fictitious dynamics of a Floquet operator that acts only on a finite region. 

Using that approach, we now indicate how a description of the fictitious dynamics in the full system can be constructed. As pointed out in Sec.~\ref{sec:density}, the density of resonances in space is small for small $\zeta$. Additionally, when $\zeta$ is small the quasienergy scales $\Omega(r)$ for different $r$ are well-separated. That is, $\Omega(r) \gg \Omega(r')$ for $r \ll r'$. Based on these observations, in this section we will argue that resonances occurring (i) in different spatial locations and (ii) on different quasienergy scales can be treated independently.

Focus for now on a particular $r$. For a given $\bm{\tilde s}$, the number of $\bm{\tilde s'}$ which differ over a region of length $r$ is $\sim 2^{r-2}$ per unit length of the chain [see Eq.~\eqref{eq:pr}]. If the quasienergies associated with these $\bm{\tilde s'}$ are uniformly distributed throughout the spectrum then we expect that, per unit length, $r$-resonances are separated by fictitious time intervals $\Delta \lambda_s(r) \sim 2^{2-r}$.

The separations $\Delta \lambda_s(r)$ must be compared with the durations of resonances $\Delta \lambda_d(r)$. As is clear from Eqs.~\eqref{eq:splitting} and \eqref{eq:tanphi}, the typical duration $\Delta \lambda_d(r) \sim \Omega(r)$. Therefore, if we focus on just one lengthscale $r$, the fraction of the fictitious time over which resonances are occurring is given by
\begin{align}
	\frac{\Delta \lambda_d(r)}{\Delta \lambda_s(r)} \sim e^{1/\zeta} e^{-(1/\zeta-1/\zeta_c)r},
\label{eq:resonancefraction}
\end{align}
in each unit length of the chain, i.e. resonances are rare in $\lambda$ for small $\zeta$. Furthermore, summing the right-hand side of Eq.~\eqref{eq:resonancefraction} over $r$, one finds a finite result for $\zeta < \zeta_c$. The fraction of the fictitious time over which a resonance is occurring is therefore finite.  

To describe the $r$-resonance in the language of Sec.~\ref{sec:pairwise}, one starts by identifying the local Floquet operator $W_{br}(\lambda')$. If a resonance occurs in the spectrum of this local operator between $\lambda'$ and $\lambda$, where ${|\lambda-\lambda'| \gtrsim \Delta \lambda_d(r)}$, we construct a fictitious time evolution operator $U_{br}(\lambda,\lambda')$ as in Sec.~\ref{sec:local}. The corresponding operator for the full system is 
\begin{align}
	U_{br}(\lambda,\lambda') \otimes \mathbbm{1}_{L-br},
\label{eq:localfictitioustimeevolution}
\end{align}
where $\mathbbm{1}_{L-br}$ is the identity operator acting on the complement of the region of length $br$ that is centred on the resonance. Even restricting to a particular $r$, for large $L$ many $r$-resonances will proceed simultaneously. However, for small $\zeta$ they are unlikely to overlap in space. For resonances that do not overlap, the fictitious time evolution operators with the form in Eq.~\eqref{eq:localfictitioustimeevolution} commute with one another.

We now explain why the different values of $r$ can be considered separately at small $\zeta$. To do so we consider the case where two resonances overlap in space but take place on different lengthscales $r$ and $r'$. First, with $r'<r$, the duration $\Delta \lambda_d(r) \ll \Delta \lambda_d(r')$. Evolving the spectrum from $\lambda'$ to $\lambda$ with $|\lambda-\lambda'| \sim \Delta \lambda_d(r)$, the fictitious time evolution operator $U_{br}(\lambda,\lambda')$ for the $r$-resonance involves a swap operation on the spectrum. However, for the $r'$-resonance we have $U_{br'}(\lambda,\lambda') \sim \mathbbm{1}_{br'}$, with equality in the limit $\zeta \to 0$. In the MBL phase we therefore expect that $U_{br}(\lambda,\lambda') \otimes \mathbbm{1}_{L-br}$ and $U_{br'}(\lambda,\lambda') \otimes \mathbbm{1}_{L-br'}$ approximately commute over the interval $\Delta \lambda_d(r)$. 
We have assumed that the character of the $r$-resonance described by $U_{br}(\lambda,\lambda')$ is not affected by the ongoing $r'$-resonance. In other words, we have assumed that the character of resonances on large lengthscales is not significantly affected by those resonances that are simultaneously occurring on small lengthscales.

Second, with $r'>r$, we have ${\Delta \lambda_d(r') \ll \Delta \lambda_d(r)}$. The $r'$-resonances here occur on large lengthscales, small energy scales, and involve many LIOM. Consider following a LIOM configuration $\bm{\tilde s}$ through the duration $\Delta \lambda_d(r)$ of an $r$-resonance, and ask how a large a fraction of this interval is taken up by the sharp $r'$-resonances involving $\bm{\tilde s}$. The effects of these resonances are clear in the lower panel of Fig.~\ref{fig:leveldynamics}, where we see a number of sharp features in the black curve. The number of spatial regions of length $r'$ that overlap with the $r$-resonance of interest is $(r+r')$, and so there are $(r+r')2^{r'-2}$ configurations $\bm{\tilde s'}$ that could be involved in an overlapping $r'$-resonance with $\bm{\tilde s}$. The typical separation in $\lambda$ between these $r'$-resonances is therefore $(r+r')^{-1}\Delta \lambda_s(r')$. Because their duration is $\Delta \lambda_d(r')$ we find that for $L \to \infty$ a fraction 
\begin{align}
	\sum_{r'=r+1}^{\infty} (r+r') \frac{\Delta \lambda_d(r')}{\Delta \lambda_s(r')} \notag
\end{align}
of the $r$-resonance is taken up by sharp resonances on small energy scales. Crucially, this fraction is small for small $\zeta$. This means that for most of the duration $\Delta \lambda_d(r)$ of a high-energy $r$-resonance we can neglect intermittent low-energy $r'$-resonances. 

The discussion in this section suggests a way to coarse-grain the fictitious dynamics. If we view the $\lambda-\theta$ plane with resolution $\Omega(r)$, then all crossings with $|\omega| \ll \Omega(r)$ appear exact. Note that the exchange of LIOM labels between crossing levels, illustrated on the left in Fig.~\ref{fig:crossing}, is essential for such a coarse-graining to be appropriate. 

\subsection{Ensemble averaging}\label{sec:averaging}
Above, we have developed a model for the local resonances that arise in individual disorder realisations as $\lambda$ is varied, with $|\lambda| \ll 1$. We have absorbed details of the initial Floquet operator $W$, and the operator $G$, into the complex numbers $z$ and the parameters $\lambda_0$ [see, for example, Eq.~\eqref{eq:splitting}]. The former encode the quasienergy separations $\omega_{br}(\lambda_0)$ at resonance, and the latter the quasienergy separations $\omega_{br}(0)$ at $\lambda=0$. Our treatment is strictly appropriate only in the case where $|\lambda_0| \gg \Omega(r)$, since in order to compute matrix elements of $G$ at $\lambda=0$ we have assumed that the relevant pairs of eigenstates are not resonant. Following an average over $W$ and $G$, however, physical quantities computed at different values of $\lambda$ have equivalent statistical properties, so we can ignore the restriction to $|\lambda_0| \gg \Omega(r)$.

In order to develop a statistical theory from our model for resonances, it is necessary to perform averages over $z$ and $\lambda_0$. We allow the values of $z$ and $\lambda_0$ to be independent for distinct local resonances, and choose distributions
\begin{align}
	p_z(z) &= [2\pi]^{-1} e^{-|z|^2/2} \label{eq:distszlambda} \\
	p_{\lambda_0}(\lambda_0) &= [2\Lambda]^{-1}, \notag 
\end{align}
where $|\lambda_0|<\Lambda$. The distribution of $z$ is arbitrary, but our results are sensitive only to the facts that $z$ is complex, random, and typically of order unity. With TRS, one should instead choose real $z$. The distribution $p_{\lambda_0}(\lambda_0)$ can be rationalised as follows. To a first approximation we expect that $\omega_{br}(0)$ is uniformly distributed on $[-\pi,\pi)$. Then, for $|\partial_{\lambda}\omega_{br}(\lambda)| \sim 1$, we have $\lambda_0 \sim \omega_{br}(0)$. Note that this implies $\Lambda \sim \pi$

\section{Spectral statistics}\label{sec:spectrum}
Using this picture of locally pairwise resonances we can determine the spectral statistics in the MBL phase. As is well-known, for large $L$ the two-point correlator of the level density is close to the Poisson form for uncorrelated levels. This is because the probability for a typical pair of levels to differ in their LIOM configuration over a finite lengthscale $r$, which would allow for a resonance on quasienergy scale $\Omega(r)$, decays with increasing $L$ as $\sim L 2^{r-L}$. Therefore, typical pairs of levels do not resonate on any finite energy scale in the thermodynamic limit $L \to \infty$. There are nevertheless $\sim L 2^{L+r}$ pairs of levels that can resonate on energy scale $\Omega(r)$. These resonances are responsible for residual level repulsion in the MBL phase.

\subsection{Two-point correlator}\label{sec:twopoint}

Here we calculate the two-point correlator of the level density, $p_{\omega}(\omega)$, defined in Eq.~\eqref{eq:pomegadef}. Our starting point for the calculation is the expression Eq.~\eqref{eq:splitting}, which gives the form of the quasienergy separations for Floquet operators that act on finite spatial regions. 

Before we discuss how to apply this expression to a large system, we determine its distribution, $p_{\omega|r}(\omega_{br},r)$, using Eq.~\eqref{eq:distszlambda}. The result is [see Appendix~\ref{app:omegadist} for details]
\begin{align}
	p_{\omega|r}(\omega_{br},r) \simeq \begin{cases} \pi^{-1} [\omega_{br}/\Omega(r)]^{\beta}, \, &|\omega_{br}| \ll \Omega(r) \\
	[2\pi]^{-1}, \, &\Omega(r) \ll |\omega_{br}| \end{cases},
\label{eq:pomegar}
\end{align}
where $\beta=2$, and we have set $\Lambda=\pi$ under the assumption that $\Omega(r) \ll 1$. Here the effect of level repulsion is manifest in the reduction of $p_{\omega|r}(\omega_{br},r)$ for $|\omega_{br}| \ll \Omega(r)$.

Turning now to the many-body spectrum, we consider a particular eigenstate $\ket{n}$ in the sum in Eq.~\eqref{eq:pomegadef}. The sum over $\ket{m}$ with $m \neq n$ can then be organised according to the kinds of resonances that $\ket{n}$ and $\ket{m}$ may be involved in. Focusing on quasienergy scale $\Omega(r)$, for example, $\ket{n}$ and $\ket{m}$ may resonate with one another over multiple regions of length $r$. Then, to determine the distribution of $(\theta_n-\theta_m)$, recall that for large $r$ or small $\zeta$ distinct $r$-resonances are typically separated in space by distances $\rho^{-1}(r) \gg r$. This means that we can treat the different resonances as independent. The contribution to $(\theta_n-\theta_m)$ on scale $\Omega(r)$ is therefore given by a sum over contributions from these resonances. We are then concerned with the statistical properties of a sum of independent random variables each having distribution Eq.~\eqref{eq:pomegar}. Crucially, although the distribution of $\omega_{br}$ is suppressed for $|\omega_{br}| \ll \Omega(r)$, the distribution of a sum of such terms is not. To determine the reduction of $p_{\omega}(\omega)$ from $[2\pi]^{-1}$ on scale $|\omega| \sim \Omega(r)$, we need only consider pairs of eigenstates $\ket{n}$ and $\ket{m}$ that resonate over a single region of length $r$.

As an example of a contribution involving more than one resonance, suppose that the spectrum of the Floquet operator features $r$-resonances in two regions $A$ and $B$, separated in space by $\rho^{-1}(r)$. In the language of Sec.~\ref{sec:local}, the local Floquet operators $W^A_{br}$ and $W^B_{br}$ feature pairwise resonances with respective quasienergy splittings $\omega^A_{br}$ and $\omega^B_{br}$. These are distributed according to Eq.~\eqref{eq:pomegar}. Taking as reference a many-body eigenstate that participates in both resonances, there are three quasienergy separation to consider. First there is $\omega \simeq \omega^A_{br}$, the separation between our reference and the state with which it resonates in region $A$ only. Second, there is $\omega \simeq \omega^B_{br}$. Level repulsion on scale $\Omega(r)$ is manifest in the distributions of these separations for $|\omega| \lesssim \Omega(r)$. The third separation is between our reference and the state with which it resonates in both $A$ and $B$, and this is $\omega \simeq \omega^A_{br}+\omega^B_{br}$. On scale $|\omega| \sim \Omega(r)$, the probability density of this quantity is approximately $[2\pi]^{-1}$. The effects of level repulsion set in only on a much smaller scale, $\Omega(\rho^{-1}(r))$, which is the strength of the coupling between $A$ and $B$.

\begin{figure}
\includegraphics[width=0.47\textwidth]{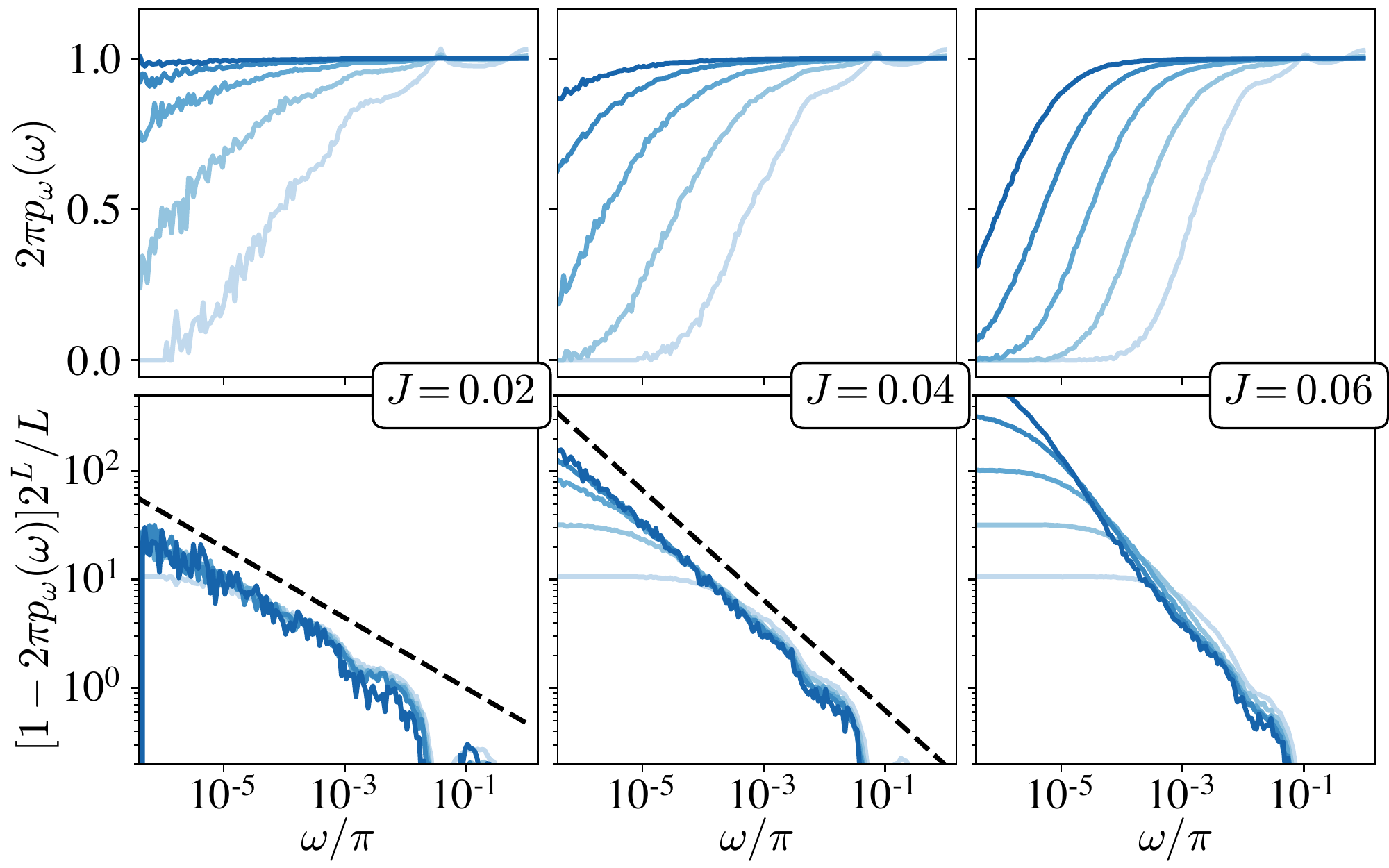}
\caption{Two-point correlator of the level density $p_{\omega}(\omega)$ [Eq.~\eqref{eq:pomegadef} for various $J$ (columns) and for $L=6,8,\ldots,14$ (from light to dark). Upper panels show $\pi p_{\omega}(\omega)$, equal to unity for uncorrelated levels, and lower panels show deviations $1-2\pi p(\omega)$ scaled by $2^L/L$. The dashed lines indicate power-law fits [see Sec.~\ref{sec:zeta} and Fig.~\ref{fig:zeta}.}
\label{fig:pomega}
\end{figure}

In summary, the effects of level repulsion on scale $\Omega(r)$ that are manifest in $p_{\omega}(\omega)$ are between pairs of many-body eigenstates that resonate with one another over a single region of length $r$. The number of many-body eigenstates $\ket{m}$ that can resonate with a given $\ket{n}$ in this way is simply $p_r(r)$ in Eq.~\eqref{eq:pr}. Using Eq.~\eqref{eq:pomegar} we can determine the overall distribution $p_{\omega}(\omega)$ via
\begin{align}
	p_{\omega}(\omega) &= \sum_{r=2}^{L} p_{\omega |r}(\omega,r)p_r(r). \label{eq:pomega1}
\end{align}
As we have discussed below Eq.~\eqref{eq:rhor}, the case $r=1$ makes a significant contribution only at high frequencies $\omega$, so we neglect it here.

We expect that Eq.~\eqref{eq:pomega1} is appropriate for arbitrarily large $L$; the function $p_r(r)$ in Eq.~\eqref{eq:pr} includes a factor $L$ (at least for $r \leq L/2$) and this accounts for the fact that $r$-resonances giving rise to level repulsion can be located anywhere in the chain. Making the approximation ${p_{\omega|r}(\omega,r) \simeq \pi^{-1} [\omega/\Omega(r)]^{\beta}}$ for all $|\omega|< 2^{-1/\beta}\Omega(r)$, and $p_{\omega|r}(\omega|r) \simeq [2\pi]^{-1}$ otherwise, we find the distribution of level separations for $e^{-L/\zeta} \ll|\omega| \ll 1$,
\begin{align}
	p_{\omega}(\omega) &= [2\pi]^{-1} \Big[ 1 - a\frac{L}{2^L}|\omega|^{-\zeta/\zeta_c} + \ldots \Big], \label{eq:pomega2}
\end{align} 
where $a$ is a constant. For small $L$ we expect a modified entropic factor arising from the different form of $p_r(r)$ at large $r$. The leading deviations from the Poisson form, which appear as the second term in Eq.~\eqref{eq:pomega2}, come from pairs of levels with $r \sim \zeta \ln|\omega|^{-1}$. The ellipses denote contributions from sub-dominant values of $r$. Note that the $\omega$-dependence of the second term in Eq.~\eqref{eq:pomega2} does not depend on $\beta$. The exponential decay with $L$ comes from the fact that typical pairs of levels have $r$ of order $L$, so repel only weakly. At finite $\omega$, their repulsion is negligible in the large $L$ limit.

The corrections to Poisson statistics in Eq.~\eqref{eq:pomega2} have the form $|\omega^*/\omega|^{\zeta/\zeta_c}$ for $|\omega| \gg \omega^*$, where
\begin{align}
	\omega^* = L^{\zeta_c/\zeta} e^{-L/\zeta}.
\label{eq:omegastar}
\end{align}
Another low-frequency regime sets in for $|\omega| < e^{-L/\zeta}$, where we expect $p_{\omega}(\omega) \sim e^{\beta L/\zeta}|\omega|^{\beta}$ arising from resonances on lengthscales $r \sim L$. Note however that for large $L$ and any $\zeta < \zeta_c$, the scale on which this second form is appropriate is much smaller than the mean level spacing $\sim 2^{-L} = e^{-L/\zeta_c}$. Similarly, $\omega^* \ll e^{-L/\zeta_c}$.

Our numerical calculations in Fig.~\ref{fig:pomega} show excellent support for Eq.~\eqref{eq:pomega2}. In the upper panels of Fig.~\ref{fig:pomega} we show $p_{\omega}(\omega)$ for various $J$ and $L$. As $J$ is increased the deviations of $p_{\omega}(\omega)$ from the Poisson result become more prominent and, for each value of $J$, increasing $L$ diminishes these deviations. In the lower panels of Fig.~\ref{fig:pomega} we investigate these deviations in more detail. In line with Eq.~\eqref{eq:pomega2} we find that on increasing $L$ at fixed $\omega$ the deviations $[1- 2\pi p_{\omega}(\omega)] \times (2^L/L)$ become approximately $L$-independent. This indicates that only resonances up to some $L$-independent value $r$ are contributing at each $\omega$. For small $L$ we expect to observe the regime where deviations are dominated by $r$-resonances with $r \sim L$, and indeed this kind of behaviour is evident in the lower panels of Fig.~\ref{fig:pomega}. For large $L$ there is a clear power-law dependence of $[1-2\pi p_{\omega}(\omega)] \times (2^L/L)$ on $\omega$, with a faster decay at larger $J$. This is exactly the behaviour expected from Eq.~\eqref{eq:pomega2}.

Turning now to the time domain, we consider the spectral form factor $K(t) \equiv |\text{Tr}W^t|^2$ defined for integer $t$ [see also Ref.~\cite{garratt2020manybody}]. The disorder average $\langle K(t)\rangle$ is related to $p_{\omega}(\omega)$ via
\begin{align}
	\langle K(t) \rangle = 2^L + 2^L(2^L-1) \int_{-\pi}^{\pi} d\omega p_{\omega}(\omega) e^{i\omega t}.
\label{eq:SFF}
\end{align}
For uncorrelated levels $p_{\omega}(\omega)$ is uniform and therefore $\langle K(t) \rangle = 2^L$. We have shown above that the deviations of $p_{\omega}(\omega)$ from a uniform distribution are suppressed by a factor $2^{-L}$, so from Eq.~\eqref{eq:SFF} we find that level repulsion in the MBL phase gives rise to a multiplicative correction to the average SFF: $\langle K(t) \rangle = 2^L[ 1 - L A(t)]$, where $A(t)$ is approximately $L$-independent and vanishes for $t \to \infty$. The average SFF therefore approaches its late time value as a power-law, $A(t) \sim t^{-(1-\zeta/\zeta_c )}$ for $t \ll t^*$ where $t^*=(2\pi)/\omega^* \sim e^{L/\zeta}$ [Eq.~\eqref{eq:omegastar}]. Then, at time $t$, $\langle K(t)\rangle$ is suppressed by the repulsion between pairs of LIOM configurations differing over lengthscales $r \sim \zeta \ln t$. On the longest timescales $t \gg t^*$ the average spectral form factor is unaffected by residual level repulsion, and $A(t) \to 0$.

\subsection{Level curvatures}\label{sec:curvatures}
In Fig.~\ref{fig:curvature1} we have shown that the distribution of level curvatures $p_{\kappa}(\kappa)$ is qualitatively different for $J=0$ and $J \neq 0$. In particular, we have shown that with $J \neq 0$ a heavy tail appears at large $\kappa$. This heavy tail arises from avoided level crossings. In this section we analytically determine the form of this tail using our model for the resonances set out in Sec.~\ref{sec:resonances}. In order to do so, we first argue that the total curvature of a level can be computed as the sum of contributions from all possible local resonances.

We start by considering a single local resonance. For a resonance on lengthscale $r$, our description is based on the spectral properties of a Floquet operator $W_{br}(\lambda)$ that acts on a finite region of $br$ sites. The level separation associated with the resonance has the form $\omega_{br}(\lambda)$ [Eq.~\eqref{eq:splitting}], and this gives a contribution to the curvature that we denote $\tilde \kappa_{br}(\lambda) \equiv \pm \frac{1}{2}\partial^2_{\lambda} \omega_{br}(\lambda)$. Explicitly,
\begin{align}
	\tilde \kappa_{br}(\lambda;\lambda_0,z,r) = \pm \frac{1}{2}\frac{|z|^2 \Omega^2(r)}{[(\lambda-\lambda_0)^2 + |z|^2 \Omega^2(r)]^{3/2}}.
\label{eq:tildekappapairwise}
\end{align}
The total contribution to the curvature of a quasienergy of $W_{br}(\lambda)$ arising from $r$-resonances can be estimated by summing $2^{r-2}$ terms of the form $\tilde \kappa_{br}(\lambda;z,r)$, allowing each term in the sum to have a different value of $z$ and centre $\lambda_0$. 

In a large system, a typical eigenstate of $W(\lambda)$ participates in multiple local resonances that are in different locations. We see from Eq.~\eqref{eq:tildekappapairwise} that $r$-resonances are associated with $\tilde \kappa_{br} \sim \Omega^{-1}(r)$, and we can ask about the various contributions to the total curvature $\kappa$ that are on this scale. Following a similar line of argument as after Eq.~\eqref{eq:pomegar}, when distinct $r$-resonances can be treated as independent we sum their contributions to the total curvature.

We model each of these contributions using Eq.~\eqref{eq:tildekappapairwise}. The resulting expression has the form
\begin{align}
	\kappa = \sum_{r=2}^L \sum_{i=1}^{2^L p_r(r)} \tilde \kappa_{br}(\lambda;\lambda_i,z_i,r),
\label{eq:kappasum1}
\end{align}
where the second sum is over all $2^L p_r(r)$ possible $r$-resonances, the parameters $\lambda_i$ and $z_i$ are independent for different $i$, with distributions in Eq.~\eqref{eq:distszlambda}, and $\tilde \kappa_{br}$ is either positive or negative with equal probability. We have neglected the case $r=1$, which gives rise to the plateau at small $\kappa$ visible in Fig.~\ref{fig:curvature1}. So that we can determine the distribution $p_{\kappa}(\kappa)$ analytically, we treat $r$ as random with distribution $p_r(r)$. This gives
\begin{align}
	\kappa \simeq  \sum_{i=1}^{2^L} \tilde \kappa(\lambda;\lambda_i,z_i,r_i),
\label{eq:kappasum2}
\end{align}
where because $r$ is now a random variable we no longer make reference to a particular local Floquet operator $W_{br}(\lambda)$, so we omit the subscript on $\tilde \kappa$, which is nevertheless given by Eq.~\eqref{eq:tildekappapairwise}. Note that, because the sum in Eq.~\eqref{eq:kappasum1} is only over $1 < r \leq L$, there is a change in the normalisation of $p_r(r)$ that is exponentially small in $L$. Similarly, we have allowed the sum in Eq.~\eqref{eq:kappasum2} to run over $2^L$ terms instead of $2^L-L-1$, which is the number of possible resonances with $r > 1$. Both of these effects are unimportant even for moderate $L$, and we neglect them. The advantage of moving to the expression Eq.~\eqref{eq:kappasum2} is that the different terms in the sum are independently and identically distributed.

Note than an alternative approach to evaluating the total curvature $\kappa_n$ of a many-body eigenstate $\ket{n}$ could start from Eq.~\eqref{eq:curvature}. This requires the calculation of matrix elements $\braket{n|G|m}$. However, at a given $\lambda$, the eigenstates $\ket{n}$ and $\ket{m}$ may resonate in multiple spatial locations. The statistical properties of $\braket{n|G|m}$ are complicated by the fact that the locations of contributing resonances are constrained by the LIOM configurations associated with $\ket{n}$ and $\ket{m}$.

Continuing from Eq.~\eqref{eq:kappasum2}, we now determine the distribution of $\tilde \kappa$. The distribution conditioned on $r$ is
\begin{align}
	p_{\tilde \kappa|r}(\tilde \kappa',r) = \int d^2 z d \lambda p_z(z) p_{\lambda_0}(\lambda_0) \delta\big[ \tilde \kappa' - \tilde \kappa_{br}(\lambda;\lambda_0,z,r)\big],\notag
\end{align}
As we discuss in Appendix~\ref{app:curvaturedist}, in the regime $1 \ll |\tilde\kappa| \ll \Omega^{-1}(r)$ one finds $p_{\tilde\kappa|r}(\kappa,r) \sim \Lambda^{-1} \Omega^{2/3}(r) |\tilde\kappa|^{-4/3}$. For $|\tilde\kappa| \gg \Omega^{-1}(r)$, on the other hand, there is a sharp decay with increasing $|\tilde \kappa|$, $p_{\tilde \kappa|r}(\tilde \kappa,r) \sim \Lambda^{-1}\Omega^{-2}(r)|\tilde \kappa|^{-4}$. The behaviour $\sim |\tilde \kappa|^{-4}$ is that expected from RMT \cite{gaspard1990parametric}, and here applies near resonances.

\begin{figure}
\includegraphics[width=0.47\textwidth]{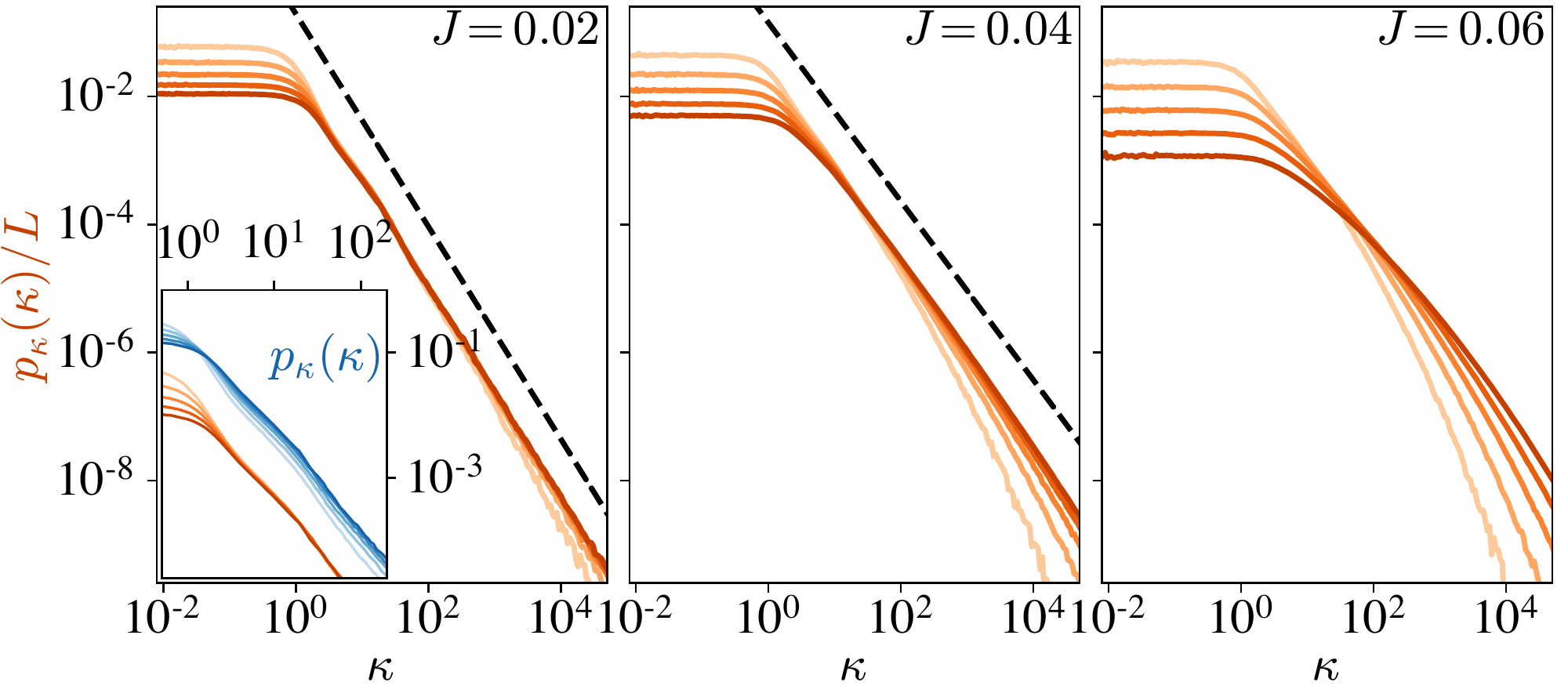}
\caption{Distributions of curvatures $p_{\kappa}(\kappa)$ for various $J$ (columns) and for $L=6,8,\ldots,14$ (from light to dark). Orange data shows $p_{\kappa}(\kappa)/L$ [see Eq.~\eqref{eq:pkappa}] and blue data in the inset shows $p_{\kappa}(\kappa)$. The dashed lines indicate power-law fits [see Sec.~\ref{sec:zeta} and Fig.~\ref{fig:zeta}].}
\label{fig:curvature2}
\end{figure}

To determine the overall distribution of $\tilde \kappa$ we evaluate
\begin{align}
	p_{\tilde \kappa}(\tilde \kappa) = \sum_{r=2}^L p_{\tilde \kappa|r}(\tilde \kappa,r)p_r(r).
\label{eq:ptildekappa1}
\end{align}
At a given $\tilde \kappa$, contributions to this sum from $r \ll \zeta \ln |\tilde \kappa|$ are small. This is because it is unlikely that $r$-resonances have $|\tilde\kappa|$ much greater than $\Omega^{-1}(r)$. On the other hand, for $r \gg \zeta \ln |\tilde \kappa|$, or equivalently $|\tilde \kappa| \ll \Omega^{-1}(r)$, we have $p_{\kappa|r}(\tilde \kappa,r) \sim \Omega^{2/3}(r) |\tilde \kappa|^{-4/3}$. The factor $\Omega^{2/3}(r)$ decays as $e^{-2r/3\zeta}$, so for $\zeta < 2\zeta_c/3$ the product $\Omega^{2/3}(r)p_r(r)$ decays with $r$. These considerations imply that for ${\zeta < 2\zeta_c/3}$, $p_{\tilde \kappa}(\tilde \kappa)$ is dominated by contributions with $r \sim \zeta \ln |\tilde \kappa|$. This leads to
\begin{align}
	p_{\tilde \kappa}(\tilde \kappa) \sim 2^{-L} L |\tilde \kappa|^{-(2-\zeta/\zeta_c)}
\label{eq:ptildekappa2}
\end{align}
at large $|\tilde \kappa|$. Different behaviour sets in when the sum in Eq.~\eqref{eq:ptildekappa1} is dominated by $r \sim L$, and so when $|\tilde \kappa| \gtrsim e^{L/\zeta}$. In the following we focus on large $L$, where we can neglect this regime.

From the distribution of $\tilde \kappa$ in Eq.~\eqref{eq:ptildekappa2}, we now determine the distribution of curvatures $\kappa$ using Eq.~\eqref{eq:kappasum2}. Note that, because $p_{\tilde \kappa}(\tilde \kappa)$ has heavy power-law tails, the second moment $\langle \tilde \kappa^2 \rangle$ does not exist, and consequently the standard central limit theorem does not apply. We start from the moment generating function for the curvature distribution,
\begin{align}
	g_{\kappa}(q) = \int_{-\infty}^{\infty} d\kappa e^{i q \kappa} p_{\kappa}(\kappa),
\end{align}
and we define $g_{\tilde \kappa}(q)$ similarly. These functions are related by $\ln g_{\kappa}(q) = 2^L \ln g_{\tilde \kappa}(q)$. From the large-$\tilde \kappa$ behaviour of $p_{\tilde \kappa}(\tilde \kappa)$, we have
\begin{align}
	1-g_{\tilde \kappa}(q) \sim 2^{-L}L|q|^{1-\zeta/\zeta_c} + \ldots
\end{align}
at small $|q|$, where the ellipses denote terms that are sub-leading in this limit. From this,
\begin{align}
	\ln g_{\kappa}(q) \sim L|q|^{1-\zeta/\zeta_c} + \ldots
\end{align}
The dependence of $g_{\kappa}(q)$ on $|q|$ in this limit is the same as that of $g_{\tilde \kappa}(q)$, but the leading term has no exponential dependence on $L$. As a result, we find
\begin{align}
	p_{\kappa}(\kappa) \sim L |\kappa|^{-(2-\zeta/\zeta_c)}
\label{eq:pkappa}
\end{align}
at large $\kappa$, i.e. unlike in $p_{\tilde \kappa}(\tilde \kappa)$, there is substantial weight in the tail of $p_{\kappa}(\kappa)$ at large $L$. The scaling $\sim L$ in Eq.~\eqref{eq:pkappa} comes from the translational freedom in the location of the resonance. The presence of this factor $L$ implies that the dominant contribution to the tail of the curvature distribution comes from pairs of states that are connected by a single resonance. These are the same pairs of states that are responsible for deviations of $p_{\omega}(\omega)$ from Poisson form, as discussed in Sec.~\ref{sec:spectrum}.

In deriving Eq.~\eqref{eq:pkappa} we neglected the contribution to $\kappa$ that comes from changes in the strengths $h_j(\lambda)$ of the local fields, and that is independent of resonances. This is the only contribution for $J=0$, and we then find that $p_{\kappa}(\kappa)$ has width of order $L^{1/2}$. Based on this we expect that Eq.~\eqref{eq:pkappa} is appropriate only for $L^{1/2} \ll |\kappa| \ll e^{L/\zeta}$. 

In Fig.~\ref{fig:curvature2} we determine $p_{\kappa}(\kappa)$ numerically, and we see that for sufficiently small $J$ and for large $|\kappa|$, the distribution decays as a power smaller than $2$. This is exactly the behaviour expected from Eq.~\eqref{eq:pkappa}, and we discuss the power of the decay in Sec.~\ref{sec:zeta}.

\section{Dynamical Correlations}\label{sec:observables}
Our model for resonances can also be applied to dynamics viewed in the frequency or time domain. From Fig.~\ref{fig:leveldynamics} it is clear that resonances have strong, and remarkably clear, signatures in the off-diagonal matrix elements of $\tau^z_j$ operators. Although resonances are rare (and so, for example, generate only a small correction to Poisson statistics), they dominate the low-frequency and long-time response of the system. In this section we first [Sec.~\ref{sec:lorentzian}] develop a theory for the statistical properties of these matrix elements, and then [Sec.~\ref{sec:specfun}] set out the consequences of this theory for the spectral functions of spin operators $\tau^{\alpha}_j$.

\subsection{Lorentzian parametric resonances}\label{sec:lorentzian}
For concreteness, consider first a single $r$-resonance with $r > 1$. To describe it we consider the local operator $W_{br}(\lambda)$ that acts on the $br$ sites centred on the resonant region, as discussed in Sec.~\ref{sec:local}. We are interested in a pair of eigenstates of $W_{br}(\lambda)$ that resonate with one another as $\lambda$ is varied. As in Sec.~\ref{sec:pairwise}, at $\lambda=0$ we denote these eigenstates by $\ket{\bm{\tilde s}}$ and $\ket{\bm{\tilde s'}}$.

If at $\lambda=0$ the spectrum of $W_{br}$ does not feature a resonance, the LIOM $\ttau^z_j$ in the region $br$ closely resemble the operators $\tau^z_j$. We therefore have $\braket{\bm{\tilde s}|\tau^z_j|\bm{\tilde s}} \simeq \pm 1$ up to corrections of order $J$. Also, the off-diagonal matrix elements of $\tau^z_j$ are small, and vanish in the limit of vanishing $J$.

However, if on varying $\lambda$ we bring our pair of eigenstates into resonance, they take the form in Eq.~\eqref{eq:plusminus}. The off-diagonal matrix elements of $\tau^z_j(\lambda)$ between our resonant pair of $W_{br}(\lambda)$ eigenstates are
\begin{align}
	&\braket{-(\lambda)|\tau^z_j(\lambda)|+(\lambda)} = \frac{1}{2} \sin\varphi(\lambda) \label{eq:offdiagz}\\ \times &\Big(\braket{\bm{\tilde s}|\tau^z_j(\lambda)|\bm{\tilde s}} - \braket{\bm{\tilde s'}|\tau^z_j(\lambda)|\bm{\tilde s'}}\Big) + \ldots \notag .
\end{align}
For $|\lambda| \ll 1$ we have $\tau^z_j(\lambda) \simeq \tau^z_j$, so that $\braket{\bm{\tilde s}|\tau^z_j(\lambda)|\bm{\tilde s}}$ and $\braket{\bm{\tilde s'}|\tau^z_j(\lambda)|\bm{\tilde s'}} \simeq \pm 1$. The ellipses in Eq.~\eqref{eq:offdiagz} denote terms of the form $\braket{\bm{\tilde s}|\tau^z_j(\lambda)|\bm{\tilde s'}}$, which we expect to be of order $J\Omega(r)$. For a site $j$ with $\bm{\tilde s}_j \neq \bm{\tilde s'}_j$, the off-diagonal matrix elements of $\tau^z_j(\lambda)$ will therefore behave as $\sin \varphi(\lambda) \sim \Omega(r)/\omega(\lambda)$ for $|\omega| \ll 1/J$. Similar considerations for $\tau^x_j(\lambda)$ suggest $\braket{-(\lambda)|\tau^x(\lambda)|+(\lambda)} \sim J\Omega(r)/\omega(\lambda)$. 

From Eqs.~\eqref{eq:offdiagz} and \eqref{eq:tanphi}, we find that the modulus-square off-diagonal matrix elements $Z_{nm,j}(\lambda)$ [Eq.~\eqref{eq:bigZeigenstates}] behave as
\begin{align}
	Z(\lambda;\lambda_0,z,r) \simeq \frac{|z|^2 \Omega^2(r)}{|z|^2 \Omega^2(r) + (\lambda-\lambda_0)^2}
\label{eq:bigZ}
\end{align}
in the vicinity of the resonance. From Eq.~\eqref{eq:offdiagz} it is clear that Eq.~\eqref{eq:bigZ} is appropriate only if the resonant LIOM configurations $\bm{\tilde s}$ and $\bm{\tilde s'}$, which correspond to the eigenstates $\ket{n}$ and $\ket{m}$, differ at the site $j$. Otherwise, $Z$ is small. We expect that the analogous quantity defined for $\tau^x_j$ should be suppressed by $\sim J^2$.


Eq.~\eqref{eq:bigZ} reveals the lineshapes of the resonances that occur as $\lambda$ is tuned: they are Lorentzian in $\lambda$ [see also the lower panel of Fig.~\ref{fig:leveldynamics}]. Starting from the distributions in Eq.~\eqref{eq:distszlambda} it is straightforward to eliminate $z$ and $\lambda_0$ and determine the joint probability distribution of $\omega$ and $Z$ for each $r$. Setting $\lambda=0$ [see Sec.~\ref{sec:averaging}] and restricting to $\omega>0$ for convenience, from Eqs.~\eqref{eq:splitting} and \eqref{eq:bigZ} we find $|z|\Omega = \omega Z^{1/2}$ and $\lambda_0 = \omega(1-Z)^{1/2}$. Therefore
\begin{align}
	p_{\omega,Z|r}(\omega,Z,r)d \omega dZ = p_{|z|\Omega}(|z|\Omega)p_{\lambda_0}(\lambda_0)d(|z|\Omega) d\lambda_0. \notag
\end{align}
Computing the Jacobian for this transformation we find
\begin{align}
	p_{\omega,Z|r}(\omega,Z,r) \sim &\Lambda^{-1}(1-Z)^{-1/2}\Big(\frac{\omega}{\Omega}\Big)^2 \notag\\ &\times \exp\Big[-\frac{1}{2}\Big(\frac{\omega}{\Omega}\Big)^2 Z\Big],
\label{eq:pomegaZr}
\end{align}
up to prefactors of order unity. The precise functional form of the decay of $p_{\omega,Z|r}(\omega,Z,r)$ at $\omega \gg \Omega(r)$ is inherited from the form we have chosen for $p_z(z)$ in Eq.~\eqref{eq:distszlambda}. However, we expect that the existence of a maximum (as a function of $\omega$) at ${\omega \sim \Omega Z^{-1/2}}$ is generic. 

To determine the distribution of $Z$ we integrate Eq.~\eqref{eq:pomegaZr} over $\omega$. Near a resonance, where $Z \gg \Omega^2(r)$, the Gaussian factor is small for the largest physical $\omega$. The integral can then be evaluated analytically, and we find
\begin{align}
	p_{Z|r}(Z,r) \sim \Lambda^{-1} \Omega(r)(1-Z)^{-1/2} Z^{-3/2} .
\label{eq:pZr}
\end{align}
Note that the factor $(1-Z)^{-1/2}Z^{-3/2}$ is simply a consequence of the Lorentzian in $\lambda$. An equivalent $3/2$ power was observed numerically in Ref.~\cite{villalonga2020eigenstates}. Other aspects of the form of Eq.~\eqref{eq:pZr} can be rationalised using Eq.~\eqref{eq:bigZ}. There we see that $Z(\lambda)$ is of order unity only for $|\lambda-\lambda_0| \lesssim \Omega(r)$, and if $\lambda_0$ is distributed uniformly over an interval $2\Lambda$, the probability for this to occur is $\sim \Lambda^{-1}\Omega(r)$. We do not expect our model for the resonances to adequately describe the off-resonant regime $Z \ll \Omega^2(r)$, but this is not our focus.

\begin{figure}
\includegraphics[width=0.47\textwidth]{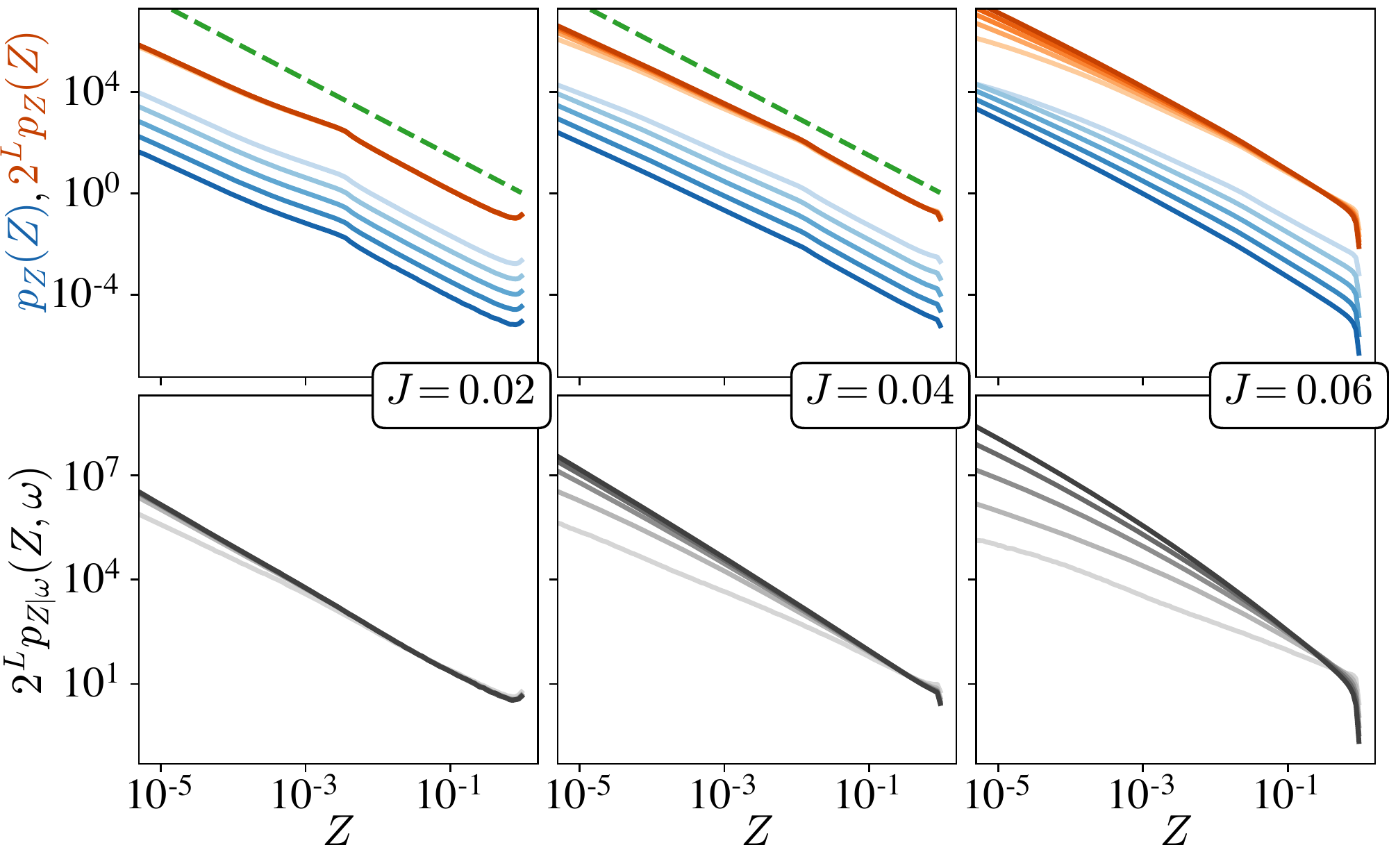}
\caption{Distributions of $Z$ for various $J$ (columns) and for $L=6,8,\ldots,14$ (from light to dark). Upper panels show the unscaled (blue) and scaled (orange) distribution, and the dashed line is a guide to eye showing decay $\propto Z^{-3/2}$. Lower panels show the scaled conditional distribution $2^L p_{Z|\omega}(\omega,Z)$ for $\omega$ in the window $10^{-4} < \omega < 10^{-3}$, testing data collapse after this scaling with $L$.}
\label{fig:offdiagdists}
\end{figure}

From Eq.~\eqref{eq:pZr} we determine the full distribution of $Z$ by summing over $r$,
\begin{align}
	p(Z) \simeq \sum_{r=2}^{L} p_{Z|r}(Z,r)p_r(r) \times \frac{r}{2L}
\label{eq:pZ}
\end{align}
where the factor $r/(2L)$ appears because the resonance must involve the site where $\tau^z_j$ acts. As usual, since we are concerned with quantities whose $r$-dependence is exponential, in the following we neglect the factor $r$ in the summand in Eq.~\eqref{eq:pZ}. Note that Eq.~\eqref{eq:pZ} is dominated by small $r$ for $\zeta < \zeta_c$. Evaluating the sum over $r$, we find
\begin{align}
	p_Z(Z) \sim 2^{-L}\Lambda^{-1}(1-Z)^{-1/2} Z^{-3/2},
\label{eq:pZ2}
\end{align}
for large $L$ and $Z<1$. Note that the system-size dependence $p_Z(Z) \sim 2^{-L}$ follows simply from the fact that $p_Z(Z)$ is dominated by finite $r$, where $p_r(r)$ is small. In the upper panels of Fig.~\ref{fig:offdiagdists} we test our prediction for $p_Z(Z)$ numerically, and find excellent agreement at large $Z$, where resonances dominate the behaviour.

We now use $p_{\omega,Z|r}(\omega,Z,r)$ in Eq.~\eqref{eq:pomegaZr} to determine the joint probability distribution $p_{\omega,Z}(\omega,Z)$. This involves summing over all values of $r$. Note that single-site resonances ($r=1$) do not contribute significantly to $p_{\omega,Z}(\omega,Z)$ for any $\omega$ or $Z$, so we neglect their contribution. As in Eq.~\eqref{eq:pZ} we have an expression
\begin{align}
	p_{\omega,Z}(\omega,Z) \simeq \sum_{r=2}^{L} p_{\omega,Z|r}(\omega,Z,r)p_r(r) \times \frac{r}{2L}.
\end{align}
Using Eq.~\eqref{eq:pomegaZr}, and again neglecting the factor $r$, this is
\begin{align}
	p_{\omega,Z}(\omega,Z) &\sim 2^{-L}\Lambda^{-1} (1-Z)^{-1/2} e^{-2/\zeta}\omega^2 \sum_{r=2}^{L/2} e^{y(r)} + \ldots \notag\\
	y(r) &\equiv \big( 2\zeta^{-1} + \ln 2 \big)r - \frac{1}{2}\Big(\frac{\omega}{\Omega(r)}\Big)^2 Z. \label{eq:pomegaZ}
\end{align}
Treated as a continuous function, $y(r)$ has a maximum at $r^* = r^*(\omega,Z)$ with
\begin{align}
	r^*(\omega,Z) = -\zeta \ln \big[ \omega Z^{1/2} \big] + \ldots
\label{eq:rstar}
\end{align}
where the ellipses denote terms that are of order unity. The dependence of $r^*(\omega,Z)$ on $\omega Z^{1/2}$ comes from ${Z \sim [\Omega(r)/\omega]^2}$ in Eq.~\eqref{eq:bigZ}. From Eq.~\eqref{eq:rstar} we can understand the different regimes of $p_{\omega,Z}(\omega,Z)$ that one can hope to observe in finite size systems as follows.

First, when $\omega Z^{1/2}$ is large, the sum in Eq.~\eqref{eq:pomegaZ} is dominated by $r \sim 2$. The functional form of $p_{\omega,Z}(\omega,Z)$ then closely resembles Eq.~\eqref{eq:pomegaZr}. Second, for ${2 \ll r^*(\omega,Z) \ll L}$, the distribution $p_{\omega,Z}(\omega,Z)$ is controlled by resonances on scales smaller than the system size, and we can expect that calculations for systems with $L$ sites capture properties of the thermodynamic limit. On the other hand, for $r^*(\omega,Z) \gtrsim L$, resonances on the scale of the system size dominate Eq.~\eqref{eq:pomegaZ}, so finite-size effects are likely to be extreme. This imposes severe limitations on numerical probes of low-$\omega$ spectra and dynamics [see, for example, Figs.~\ref{fig:pomega} and \ref{fig:specfun}].

In the regime $2 \ll r^*(\omega,Z) \ll L$ we can make analytic progress by approximating $\sum_{r=2}^{L} e^{y(r)} \simeq e^{y(r_*)}$. The result is
\begin{align}
	p_{\omega,Z}(\omega,Z) \sim &2^{-L} \Lambda^{-1} |\omega|^{-\zeta \ln 2}\label{eq:pomegaZ2} \\ &\times (1-Z)^{-1/2} Z^{-1-(\zeta/2)\ln2} \notag.
\end{align}
The dependence on $Z$ here is clearly distinct from that in Eq.~\eqref{eq:pZ2}. Note also that, because $p_{\omega}(\omega)$ is approximately constant for the values $|\omega| \gg \omega^*$ of interest, the conditional distribution $p_{Z|\omega}(\omega,Z) \sim p_{\omega,Z}(\omega,Z)$. An important feature of Eq.~\eqref{eq:pomegaZ2} is the exponential dependence on system size, and for sufficiently large $Z$ this is evident in the lower panels in Fig.~\ref{fig:offdiagdists}; there we show the conditional distribution $p_{Z|\omega}(\omega,Z)$ versus $Z$ for a particular window of $\omega$.

For small $Z$, however, the lengthscale $r^*(\omega,Z)$ may become comparable to or exceed the system size. The dependence of $p_{\omega,Z}(\omega,Z)$ then changes relative to Eq.~\eqref{eq:pomegaZ2}. In particular, when $r^*(\omega,Z) \gtrsim L$, we no longer expect $p_{\omega,Z}(\omega,Z) \sim 2^{-L}$. We indeed find that, on decreasing $Z$ (or increasing $J$) in the lower panels of Fig.~\ref{fig:offdiagdists}, the scaled distributions $2^L p_{Z|\omega}(\omega,Z)$ no longer collapse for different $L$. This effect is much more dramatic in the lower panels than in the upper ones, because in the lower panels we condition on small $\omega$, which amounts to selecting for resonances on larger lengthscales.

The above suggests a useful probe of the character of resonances in finite-size systems. We suppose that, in a given disorder realisation, we select a pair of levels and calculate the corresponding values $\omega$ and $Z$. If we find $Z$ of order unity, this indicates a resonance that is in some sense `nearby' in the ensemble of disorder realisations. Moreover, the deviations of $Z$ from unity indicate `how far' our level pair is from the middle of the resonance. From $r^*(\omega,Z)$ in Eq.~\eqref{eq:rstar} we have an estimate for the lengthscale of the resonance, and we can compare this with the system size $L$.

\subsection{Spectral functions}\label{sec:specfun}
\begin{figure}
    \centering
    \includegraphics[width=0.47\textwidth]{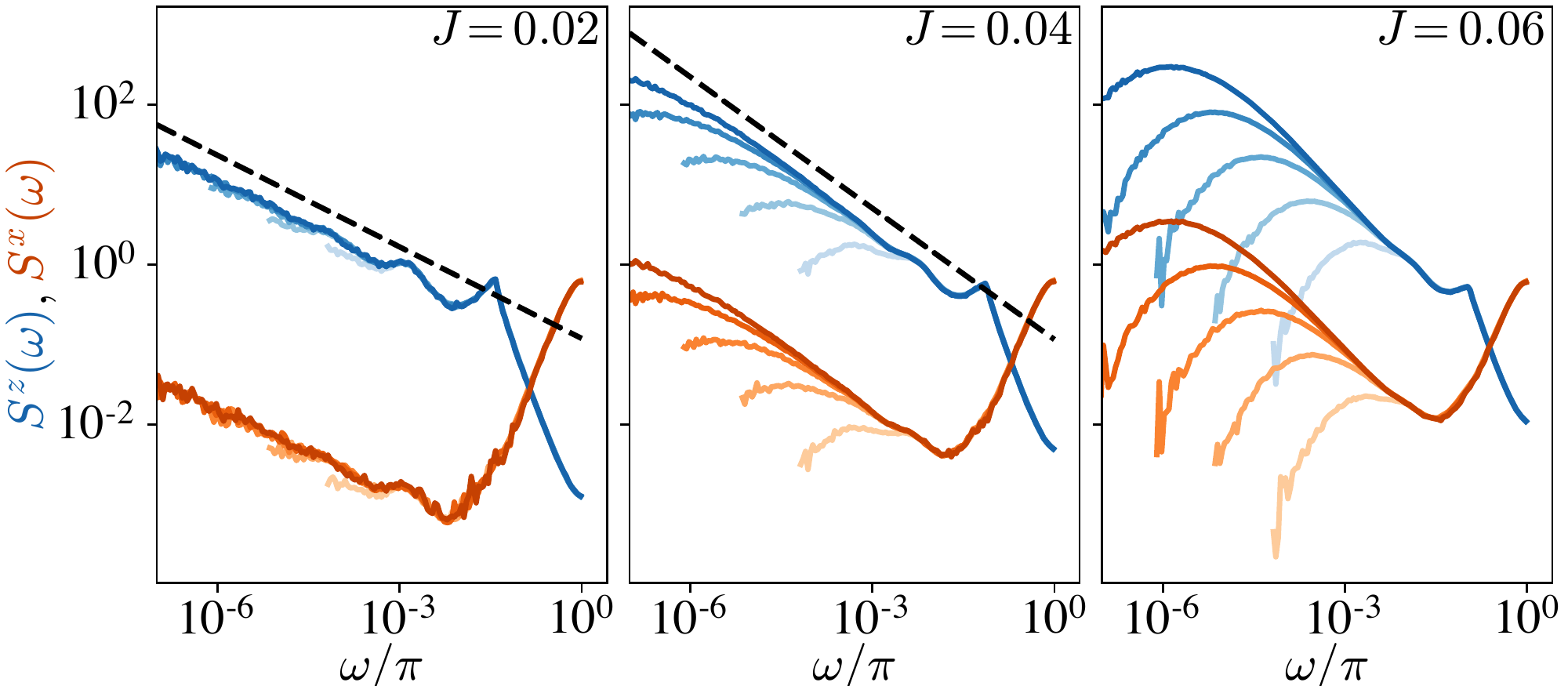}
    \caption{Spectral functions $S(\omega)$ for various $J$ and for $L=6,8,\ldots,14$ (from light to dark) for $\tau^z_j$ (blue) and $\tau^x_j$ (orange). The dashed lines indicate power-law fits [see Sec.~\ref{sec:zeta} and Fig.~\ref{fig:zeta}]. We average over all sites and over disorder.}
    \label{fig:specfun}
\end{figure}

The slow power-law decays in Fig.~\ref{fig:offdiagdists} have strong implications for the dynamics of local observables. In particular, they suggest that autocorrelation functions of spin operators are dominated by the resonances, and so by pairs of levels with large $Z$. To study the relaxation of the operators $\tau^{\alpha}_j$, we consider the spectral functions $S^{\alpha}_j(\omega)$ defined in Eq.~\eqref{eq:specfundef}. Note that these local spectral functions are not self-averaging \cite{serbyn2017thouless}, as indicated by the broad distributions in the lower panels of Fig.~\ref{fig:specfun}. In the following we focus on the disorder average $\langle S^{\alpha}_j(\omega)\rangle$.

We can infer the low-$\omega$ behaviour of $\langle S^z(\omega)\rangle$ for a given site from Eq.~\eqref{eq:pomegaZ2}, using
\begin{align}
	\langle S^z(\omega) \rangle =  2^L \int_{0}^1 dZ p_{\omega,Z}(\omega,Z)Z.
\label{eq:specfun_Zdist}
\end{align}
Due to the slow decay of the right-hand side of Eq.~\eqref{eq:pomegaZ2} with increasing $Z$, the quantity $\langle S^z(\omega)\rangle$ is dominated by contributions from resonances. Evaluating the integral in Eq.~\eqref{eq:specfun_Zdist} we find $\langle S^z(\omega) \rangle \sim |\omega|^{-\zeta/\zeta_c}$. Strikingly, this is the same power as that governing deviations of $p_{\omega}(\omega)$ from Poisson statistics [Eq.~\eqref{eq:pomega2}]. We return to this below, and also in Sec.~\ref{sec:zeta}.

To understand the result $S^z(\omega) \sim |\omega|^{-\zeta/\zeta_c}$ in more detail, we consider the contributions from different values of $r$. To this end we write
\begin{align}
	\langle S^z(\omega)\rangle = \sum_r 2^{r-2} \langle S^z_r(\omega)\rangle,
\label{eq:specfun_sumr}
\end{align}
where $\langle S^z_r(\omega)\rangle$ represents the contribution of $r$-resonances,
\begin{align}
	\langle S^z_r(\omega)\rangle = \int_0^1 dZ  p_{\omega,Z|r}(\omega,Z,r) Z.
\label{eq:specfun_sumr_intZ}
\end{align}
First we consider the regime $\omega \ll \Omega(r)$. There the exponentially-decaying factor in Eq.~\eqref{eq:pomegaZr} is approximately constant, and as a result $\langle S^z_r(\omega)\rangle \sim [\omega/\Omega(r)]^2$. On the other hand, for $\omega \gg \Omega(r)$ we need only consider $Z \ll 1$. Then $(1-Z)^{1/2} \simeq 1$ and the integral over $Z$ in Eq.~\eqref{eq:specfun_sumr_intZ} can be evaluated analytically. The result is $\langle S^z_r(\omega)\rangle \sim [\Omega(r)/\omega]^2$ for $|\omega| \gg \Omega(r)$. 

From the behaviour in these two regimes we can determine $\langle S^z(\omega)\rangle$ using Eq.~\eqref{eq:specfun_sumr}. For $r \ll \zeta \ln|\omega|^{-1}$ we have $\langle S^z_r(\omega)\rangle \sim e^{2r/\zeta}$, so the summand in Eq.~\eqref{eq:specfun_sumr} increases exponentially with $r$. For $r \gg \zeta \ln|\omega|^{-1}$ we instead find $\langle S^z(\omega)\rangle \sim e^{-2r/\zeta}$. Because $\zeta < \zeta_c$, in this regime the summand in Eq.~\eqref{eq:specfun_sumr} decreases exponentially with $r$. Therefore, the sum is dominated by $r \sim \zeta \ln|\omega|^{-1}$.

This leads to the power-law decay $\langle S^z(\omega) \rangle \sim |\omega|^{-\zeta/\zeta_c}$ at small $\omega$: the probability for a resonance with $r \sim \zeta \ln|\omega|^{-1}$ increases as $2^r \sim |\omega|^{-\zeta/\zeta_c}$, and because resonances correspond to $Z \sim 1$, we find $\langle S^z(\omega)\rangle \sim |\omega|^{-\zeta/\zeta_c}$. At large $\omega$, on the other hand, there are no many-body resonances, and $\langle S^z(\omega)\rangle$ must decrease sharply.

The other components of $\vec{\tau}_j$ have parametrically smaller autocorrelation functions at late times: based on the discussion in Sec.~\ref{sec:lorentzian}, we anticipate $\langle S^{x}(\omega)\rangle \sim J^2 |\omega|^{-\zeta/\zeta_c}$ at small $\omega$. At large $\omega$ single-site resonances contribute to these spectral functions, so their behaviour reflects that in the decoupled system: $\langle S^{x}(\omega)\rangle \sim \omega^2$. The form of this increase is inherited from the distribution of local fields $h_j$, which is model-dependent. 

We present numerical results for the spectral functions $\langle S^{x}(\omega)\rangle$ and $\langle S^{z}(\omega)\rangle$ in Fig.~\ref{fig:specfun} for various $J$, and find excellent agreement with the predictions outlined above. In particular, $\langle S^x(\omega)\rangle$ and $\langle S^z(\omega)\rangle$ decay with the same power at small $\omega$. We also find that $\langle S^x(\omega)\rangle$ and $\langle S^z(\omega)\rangle$ collapse at small $\omega$ for all $L$ when the former is scaled by $\sim (2J)^{-2}$ [not shown]. To understand the finite-size effects in Fig.~\ref{fig:specfun}, recall that $\langle S^{\alpha}(\omega)\rangle$ is dominated by resonances with $r \sim \zeta \ln|\omega|^{-1}$. For sufficiently small $\omega$, this exceeds the system size; we expect that for $r \gtrsim L$ and so $\omega \lesssim e^{-L/\zeta}$, $\langle S^{\alpha}(\omega)\rangle$ should deviate from its large-$L$ form. This behaviour is evident in our numerical results.

From the scaling of the spectral functions with $\omega$, one arrives at a power-law decay of the autocorrelation function with increasing time, $\langle C^{\alpha}(t)-C^{\alpha}(\infty)\rangle \sim t^{\zeta/\zeta_c-1}$. For larger $J$ and hence larger $\zeta$, the decay of $\langle S^{\alpha}(\omega)\rangle$ with $\omega$ at low frequencies is clearly faster, but this implies a slower approach of the autocorrelation function to its late time value. Additionally, because the same power law $|\omega|^{-\zeta/\zeta_c}$ governs deviations of $p_{\omega}(\omega)$ from Poisson statistics, both $C^{\alpha}(t)$ and the spectral form factor $K(t)$ approach their late-time values as $t^{\zeta/\zeta_c -1}$. Note that the same power law appears in the two settings because both level repulsion and dynamics on scale $\omega$ are controlled by $r$-resonances with $r \sim \zeta \ln|\omega|^{-1}$.

\section{Decay length $\zeta$}\label{sec:zeta}

A fundamental assumption of our theory is that, for disordered spin chains in the MBL phase, there exist LIOM whose support decays exponentially in space from the individual sites of the chain. In particular, if we try to construct LIOM $\tilde \tau^z_j$ in perturbation theory, starting from the operators $\tau^z_j$ which are the LIOM at $J=0$, we find that $\tilde \tau^z_j$ has support on sites $j \pm p$ at order $J^p$ \cite{ros2015integrals}. As we have discussed in connection with Eq.~\eqref{eq:LIOMexpansion}, local operators $\tau^{\alpha}_j$ can similarly be expressed in terms of $\tilde \tau^{\beta}_k$. For this reason, one expects that the off-diagonal matrix elements $\braket{n|G|m}$ behave as indicated in Eq.~\eqref{eq:bigOmega}.

In reality, $\braket{n|G|m}$ depends on details of the disorder realisation, and we have modelled this effect through the random variable $z$ [see Eq.~\eqref{eq:distszlambda}]. The effective decay length $\zeta$ in Eq.~\eqref{eq:bigOmega} can then be viewed as a parametrisation of the distribution of $\braket{n|G|m}$. At this level of approximation, physical properties at a particular $J$ are characterised by a single lengthscale $\zeta=\zeta(J)$. As indicated above, from perturbation theory we expect $e^{-p/\zeta} \sim J^p$, or
\begin{align}
	\zeta(J) = \frac{1}{\ln[J_0/J]},
\label{eq:zetaJ}
\end{align}
where $J_0$ is a constant. We expect this behaviour to apply deep within the MBL phase, with $J \ll J_c$.

A more refined picture of the MBL phase includes not only LIOM but also resonances between LIOM configuration, and we have argued that they appear at any finite $\zeta$. By considering variations of the disorder realisation, we have developed a theory for these resonances that is based only on properties of a non-resonant background of LIOM. Since this background is characterised by $\zeta$ only, so are properties of the resonances. 

A dramatic signature of the resonances is clear in the distribution of level curvatures $p_{\kappa}(\kappa)$ in Fig.~\ref{fig:curvature1}. For $J=0$ there are no large values of $\kappa$, but a heavy tail suddenly develops as soon as $J \neq 0$. In Sec.~\ref{sec:curvatures} we have calculated analytically the form of this tail, $p_{\kappa}(\kappa) \sim |\kappa|^{-(2-\zeta/\zeta_c)}$, and in this way we have related the statistical properties of resonances to the structure of our background of non-resonant LIOM. Additionally, in Sec.~\ref{sec:twopoint} we have shown how resonances give rise to deviations of $p_{\omega}(\omega)$ from Poisson statistics, and that these behave as $|\omega|^{-\zeta/\zeta_c}$ at small $\omega$. The same power-law appears in $\langle S^{\alpha}(\omega) \rangle \sim |\omega|^{-\zeta/\zeta_c}$ [see Sec.~\ref{sec:specfun}].

\begin{figure}
	\includegraphics[width=0.47\textwidth]{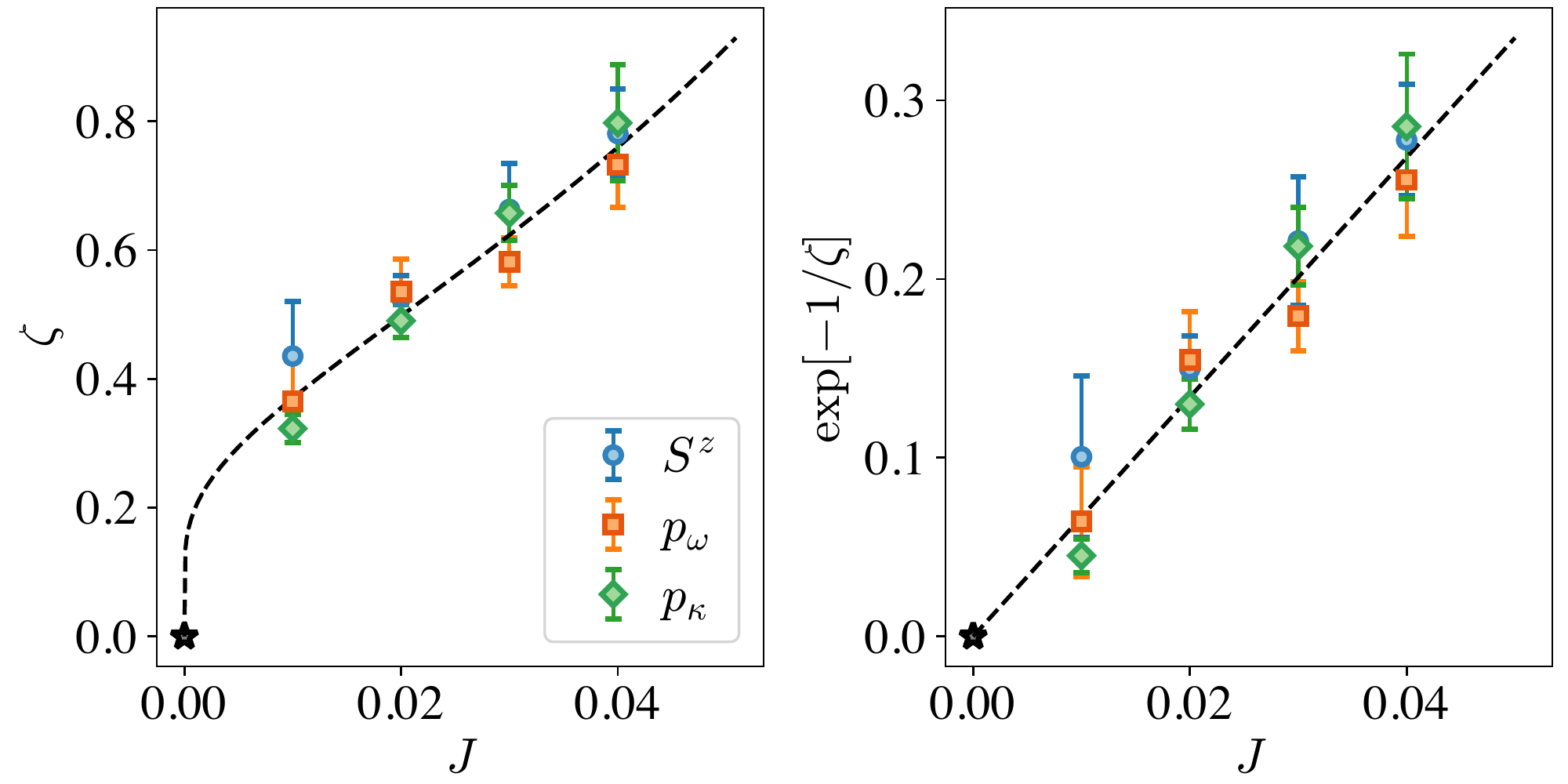}
	\caption{Decay length $\zeta(J)$ for small $J$. Blue points come from power-law fits to $\langle S^z(\omega)\rangle$ at small $\omega$ [Fig.~\ref{fig:specfun}], orange to $[1-2\pi p_{\omega}(\omega)]$ at small $\omega$ [Fig.~\ref{fig:pomega}] and green to $p_{\kappa}(\kappa)$ at large $\kappa$ [Fig.~\ref{fig:curvature2}]. Errors are dominated by systematic effects that we have estimated by varying the ranges of the fits. The dashed black line shows Eq.~\eqref{eq:zetaJ} with $J_0 \approx 0.15$, which we extract from a linear fit to $e^{-1/\zeta}$ versus $J$.}
\label{fig:zeta}
\end{figure}

For our theory to be internally consistent, the decay lengths $\zeta(J)$ that are implied by (i) the large-$\kappa$ behaviour of $p_{\kappa}(\kappa)$, and the small-$\omega$ behaviour of both (ii) $p_{\omega}(\omega)$ and (iii) $\langle S^{\alpha}(\omega)\rangle$, must agree. To test this, we extract the respective values of $\zeta(J)$ from our numerical calculations of these quantities, which are shown for certain values of $J$ in (i) Fig.~\ref{fig:curvature2}, (ii) Fig.~\ref{fig:pomega} and (iii) Fig.~\ref{fig:specfun}. The results are shown in Fig.~\ref{fig:zeta}, and we indeed find agreement between the different values of $\zeta(J)$. 

On the left in Fig.~\ref{fig:zeta} we see that $\zeta$ increases sharply from zero at small $J$, as suggested by the non-analytic behaviour in Eq.~\eqref{eq:zetaJ}. The values of $\zeta$ are significantly smaller than $\zeta_c \approx 1.4$, as required by our theory. On the right we show that $e^{-1/\zeta}$ increases approximately linearly with $J$; this is to be expected if the non-resonant LIOM for $J \neq 0$ can be constructed perturbatively from those at $J=0$. Note that we have restricted ourselves to $J \leq 0.04$ in Fig.~\ref{fig:zeta}. This is because, at larger $J$, the data in Figs.~\ref{fig:pomega}, \ref{fig:curvature2} and \ref{fig:specfun} has not converged with system size.

Above, we have inferred $\zeta$ from a comparison of our theory with numerical calculations of physical quantities. There are a variety of complementary methods that can estimate $\zeta$ via explicit construction of LIOM \cite{ros2015integrals,chandran2015constructing,pekker2017fixed,rademaker2016explicit}, and it would be interesting to compare the results of these methods with ours.

\section{Discussion}\label{sec:discussion}

In this paper we have developed a theory for resonances between LIOM configurations in the MBL phase of disordered quantum spin chains. Our approach is rooted in a fictitious dynamics of the spectral properties, and we induce this by varying the disorder realisation. The avoided level crossings that arise correspond to resonances between LIOM configurations. 
These resonances are evident in, for example, the statistics of level curvatures, and we have determined the form of the tail in the curvature distribution. Using our theory we have shown how the level repulsion associated with resonances enters the two-point correlator of the level density, and gives rise to deviations from Poisson statistics. Additionally, we have shown that the dynamical response of the MBL phase at low frequencies is dominated by resonances.

We believe our theory is appropriate deep within the MBL phase and in arbitrarily large systems. Its construction relies on the identification of evolution operators $W_{br}(\lambda)$ that act on finite spatial regions, and an approximate tensor-product decomposition of the full evolution operator $W(\lambda)$. The structure of this decomposition depends on the (quasi)energy scale of interest. We have argued that resonances can be identified with avoided crossings in the spectra of the operators $W_{br}(\lambda)$ that arise under variations in $\lambda$, and that these avoided crossings are pairwise. That is, they can be understood by considering only pairs of eigenstates of the operator $W_{br}(\lambda)$. This motivates our use of a Landau-Zener model to describe the resonances, which allows for analytic progress.


Our work should be compared to another recent approach
to describing resonances in the MBL phase and
the critical regime~\cite{crowley2021constructive}. This is focused in part on behaviour in the small systems that are accessible numerically, and in that setting the authors identify resonances with superpositions of eigenstates of the full evolution operator $W$. More generally, they argue that resonances are stable for arbitrarily large $L$ provided the disorder is strong. In our language, resonances in small systems occur at the level of the evolution operator $W$ for the full system, while for larger systems we consider separately the evolution operator $W_{br}$ for the resonant region. This has the advantage of displaying eigenstates explicitly as tensor products of factors representing local resonances and non-resonant regions.

The construction based on local operators makes clear how resonances, and hence avoided crossings in the fictitious dynamics, can occur between levels that are not neighbours in the many-body spectrum. In fact, resonances between nearest neighbours are atypical in terms of their influence on properties of the system at $L$-independent values of $\omega$. For a resonance on lengthscale $r$, the avoided crossing has a (quasi)energy width of order $\Omega(r)$. At large $L$, this greatly exceeds the mean level spacing. The implication is that pairs of levels that resonate on finite (quasi)energy scales are separated in the spectrum by a number of levels that grows exponentially with $L$.

It is interesting to ask how our picture would change in the presence of a conservation law. The introduction of a conserved energy or particle number density can be viewed as imposing a constraint on the resonances that can occur \cite{crowley2021constructive}, and one manifestation of this constraint is a reduction of the factor $p_r(r)$ in Eq.~\eqref{eq:pr}. A further question concerns the effect of TRS. To introduce TRS one can simply choose the parameter $z$ [see e.g. Eq.~\eqref{eq:splitting}] to be real, and this changes the statistical properties of individual avoided crossings. For example, in Eq.~\eqref{eq:pomegar} one would instead find the level repulsion exponent ${\beta=1}$. However, the $\omega$-dependence of $p_{\omega}(\omega)$ and $\langle S^z(\omega) \rangle$, as well as the power of the tail in $p_{\kappa}(\kappa)$, are independent of $\beta$.

The idea of using fictitious level dynamics to discuss many-body localisation was introduced in Ref.~\cite{serbyn2016spectral}. Our viewpoint and results differ in a number of important ways from that work. Most significantly, our considerations centre on resonances, which there play no explicit role. Additionally, level repulsion is in Ref.~\cite{serbyn2016spectral} described within a mean-field approximation, with the strength of the repulsion allowed to depend only on the energy separation between the states. Within our approach a much richer description emerges for the MBL phase, in which the repulsion between levels depends not only on their (quasi)energy separation, but also on the spatial structure of the associated LIOM configurations. A further distinction from our work is that Ref.~\cite{serbyn2016spectral} considers Brownian motion through the disorder ensemble, as opposed to smooth paths, and this masks the connection between avoided crossings and resonances.

Throughout, we have restricted ourselves to a characterisation of the MBL phase based on $\zeta$ alone. This clearly breaks down at $\zeta=\zeta_c$, although the true MBL transition may lie at a smaller $\zeta$ [see Ref.~\cite{morningstar2021avalanches} for a recent discussion]. Within a description based on $\zeta$, we can nevertheless ask which aspects of our treatment fail as $\zeta_c$ is approached. Focusing first on resonances that occur on a particular lengthscale $r$, increasing $\zeta$ causes the fraction of length of the system involved in $r$-resonances to grow as $r\rho(r)$. For small $r$ this may exceed unity for $\zeta$ well below $\zeta_c$; for these high-energy resonances one would be forced to ask what happens when they overlap in space. For larger $r$, however, the fraction $r \rho(r)$ remains below unity until a larger value of $\zeta$, closer to $\zeta_c$. This suggests that our description of resonances on the lowest (quasi)energy scales remains appropriate even for relatively large $\zeta$. There is, however, an open question of how they are affected by interactions between spatially-overlapping high-energy resonances that occur on small lengthscales.

As this paper was being finalised Ref.~\cite{morningstar2021avalanches} appeared, which focuses on the regime of system-wide resonances that occur between levels that are nearby on the scale of the many-body level spacing. That regime is complementary to the one that we consider.

\begin{acknowledgements}
We are grateful to A. Chandran, D. A. Huse, M. Fava, D. E. Logan and S. A. Parameswaran for useful discussions. This work was supported in part by EPSRC Grants No. EP/N01930X/1 and EP/S020527/1.
\end{acknowledgements}

\appendix

\section{Fictitious time evolution operator}\label{app:perturb}

In this appendix we recall basic aspects of perturbation theory for the spectral properties of unitary operators. Suppose $W$ is a unitary matrix with $W\ket{n} = e^{i\theta_n}\ket{n}$, and define $W(\lambda) = e^{i\lambda G}W$, where $G$ is a Hermitian matrix and $\lambda$ is small. Writing $W(\lambda)\ket{n(\lambda)}=e^{i\theta_n(\lambda)}\ket{n(\lambda)}$, we find
\begin{align}
\theta_n(\lambda) &= \theta_n + \lambda \theta^{(1)}_n + \ldots \\
\ket{n(\lambda)} &= \ket{n} + \lambda \sum_{m \neq n} C_m^{(1)} \ket{m} + \ldots \notag
\end{align}
where ellipses denote terms that are of order $\lambda^2$ and higher. Matching terms by powers of $\lambda$ in $W(\lambda)\ket{n(\lambda)}=e^{i\theta_n(\lambda)}\ket{n(\lambda)}$, we find at first order
\begin{align}
	\theta_n^{(1)} &= \braket{n|G|n} \\
	C_m^{(1)} &= -\frac{i\braket{m|G|n}}{e^{i[\theta_m-\theta_n]}-1}. \notag
\end{align}
At second order we find, for the quasienergy shifts,
\begin{align}
	\theta^{(2)}_n = \frac{1}{2}\sum_{m \neq n} |\braket{m|G|n}|^2 \cot[(\theta_n-\theta_m)/2], \notag
\end{align}
and we have used this to determine $\kappa_n$ in Eq.~\eqref{eq:curvature}.

Quite generally, we can define a fictitious time evolution operator for the eigenstates of $W(\lambda)$. Its matrix elements are
\begin{align}
	[U(\lambda,\lambda')]_{nm} = \braket{n(\lambda)|m(\lambda')}.
\end{align}
For $W(\lambda) = e^{i\lambda G}W$ we have the evolution equations
\begin{align}
	\partial_{\lambda} \theta_n &= \braket{n(\lambda)|G|n(\lambda)}\notag \\
	\partial_{\lambda} U(\lambda,\lambda') &= M(\lambda)U(\lambda,\lambda'),
\label{eq:Mfic}
\end{align}
where from unitary perturbation theory,
\begin{align}
	[M(\lambda)]_{nm} = i \frac{\braket{n(\lambda)|G|m(\lambda)}}{e^{i[\theta_n(\lambda)-\theta_m(\lambda)]}-1},
\label{eq:Mfictitious}
\end{align}
for $n \neq m$, while $[M(\lambda)]_{nn}=0$.

\section{Avoiding crossing of two levels}\label{app:pairwise}

Here we apply the framework described at the end of Appendix~\ref{app:perturb} to the spectral properties of the operators $W_{br}(\lambda)$ introduced in Sec.~\ref{sec:local}. We have argued in Sec.~\ref{sec:pairwise} that it suffices to consider the fictitious dynamics of a pair of levels of the operator $W_{br}(\lambda)$. For this reason we now discuss the solution of the above equations for two levels. 

We choose the fictitious time $\lambda=0$ as a reference point, and work in a basis defined by two eigenstates of $W_{br}$. Because resonances are rare in $\lambda$ for small $J$, with high probability our basis states do not participate in an $r$-resonance. For this reason we label them as standard LIOM configurations $\ket{\bm{\tilde s}}$ and $\ket{\bm{\tilde s'}}$. To parametrise variations in the eigenstates with $\lambda$ we introduce the coordinate $\varphi(\lambda)$ as in Eq.~\eqref{eq:plusminus}, which we repeat here for completeness
\begin{align}
\ket{+(\lambda)} &= \cos[\varphi(\lambda)/2]\ket{\bm{\tilde s}}+\sin[\varphi(\lambda)/2]\ket{\bm{\tilde s'}} \notag \\
\ket{-(\lambda)} &= \cos[\varphi(\lambda)/2]\ket{\bm{\tilde s'}}
-\sin[\varphi(\lambda)/2]\ket{\bm{\tilde s}}. \notag
\end{align}

The behaviour of $\varphi(\lambda)$ is determined by the matrix $G_{br}$. We parametrise $G_{br}$ in the basis defined by $\ket{\bm{\tilde{s}}}$ and $\ket{\bm{\tilde s'}}$ in terms of real coefficients $g_0$, $g_1$, $g_2$ and $g_3$, as
\begin{align}
	G_{br} = \begin{pmatrix}g_0 + g_3 & g_1 -ig_2\\g_1 + ig_2& g_0 - g_3\end{pmatrix}.
\end{align}
Without loss of generality, we can pick the relative phase of $\ket{\bm{\tilde s}}$ and $\ket{\bm{\tilde s'}}$ so that $g_2=0$, and we make this choice in the following. If the states $\ket{\bm{\tilde s}}$ and $\ket{\bm{\tilde s'}}$ are far from a resonance, in the sense that their quasienergy separation $\omega(\lambda=0) \gg \Omega(r)$, we expect the matrix elements of $G_{br}$ to behave as discussed in Sec.~\ref{sec:local}. That is, $g_3 \sim 1$ and $g_1 \sim \Omega(r)$.

Since we are interested in resonances occurring on scales $\Omega(r) \ll 1$, we approximate the denominator in Eq.~\eqref{eq:Mfictitious} at first order in the quasienergy difference $\omega(\lambda)=\theta_+(\lambda)-\theta_-(\lambda)$ in the exponent. Then Eq.~\eqref{eq:Mfic} reduces to
\begin{align}
\partial_\lambda \omega(\lambda) &= 2[g_3 \cos \varphi(\lambda) + g_1 \sin\varphi(\lambda)]\notag \\
\omega(\lambda) \partial_\lambda \varphi(\lambda) &= 2[g_1 \cos\varphi(\lambda) - g_3 \sin\varphi(\lambda)],
\end{align}
with the boundary conditions $\omega(0) = \theta_+(0) - \theta_-(0)$ and $\varphi(0) = 0$. 

These equations have the solution 
\begin{align}
\omega(\lambda) &= \omega(0) \frac{\sin[\varphi_0]}{\sin[\varphi_0-\varphi(\lambda)]}\,,
\end{align}
where $\tan\varphi_0\equiv g_1/g_3$, and 
\begin{align}
\cot[\varphi_0-\varphi(\lambda)] &= \frac{g_3(\lambda-\lambda_0)}{\omega(0)\sin[\varphi_0]}\,,
\label{eq:cotphi}
\end{align}
with $\lambda_0$ an integration constant. This solution describes passage through a resonance, since (taking $g_1$ and $g_3>0$ for definiteness) as $\lambda$ increases from $-\infty$ to $\infty$, $\varphi(\lambda)$ increases from $\varphi_0$ to $\varphi_0+\pi$, with a minimum in $\omega(\lambda)$ at $\lambda=\lambda_0$, where $\varphi(\lambda_0) = \varphi_0 + \pi/2$: the location of the resonance centre. Note that this increase in $\varphi(\lambda)$ by $\pi$ implies exchange of the eigenstates in Eq.~\eqref{eq:plusminus}. The expression for the quasienergy splitting can conveniently be rewritten in the standard Landau-Zener form, as
\begin{align}
	\omega(\lambda) = \sqrt{\omega^2(\lambda_0) + g_3^2 (\lambda - \lambda_0)^2}\,.
\end{align}
In addition, Eq.~\eqref{eq:cotphi} gives
\begin{align}
	\omega(\lambda_0) = \omega(0) \sqrt{\frac{g_1^2}{g_1^2+g_3^2}}.
\end{align}

The results in this appendix motivate the statistical model for a resonance defined by Eqs.~\eqref{eq:splitting}, \eqref{eq:tanphi}, and \eqref{eq:distszlambda}. There we set $g_3=1$, and $g_1=|z|\Omega(r)/\omega(0)$, so that with $|z|\Omega(r) \ll 1$ we find $\omega(\lambda_0) \simeq |z|\Omega(r)$. Then, in the vicinity of the resonance, $|\varphi(\lambda)| \gg |\varphi_0|$, and Eq.~\eqref{eq:cotphi} implies Eq.~\eqref{eq:tanphi}.

\section{Two-point correlator of the level density}\label{app:omegadist}
Here we discuss additional aspects of the calculation of $p_{\omega}(\omega)$. As in Sec.~\ref{sec:twopoint} we start with the distribution of $\omega_{br}$ [Eq.~\eqref{eq:splitting}]. This is given by 
\begin{align}
	p_{\omega|r}(\omega',r) &= \int d^2 z p_z(z) d\lambda p_{\lambda_0}(\lambda_0) \delta\big[ \omega' - \omega_{br}(\lambda;\lambda_0,z,r)\big], \notag
\end{align}
where $p_z(z)$ and $p_{\lambda_0}(\lambda_0)$ are given in Eq.~\eqref{eq:distszlambda}, and we are free to choose $\lambda=0$. Fixing the cutoff on the $\lambda_0$ distribution to $\Lambda=\pi$ we find, for ${\omega'_{br} \gg \Omega(r)}$,
\begin{align}
	p_{\omega|r}(\omega',r) &\simeq [2\pi]^{-1} \int d\lambda_0 \delta[\omega' - \lambda_0] = [2\pi]^{-1}
\label{eq:pomegarappendix1}
\end{align}
To determine $p_{\omega|r}(\omega',r)$ for $|\omega'| \ll \Omega(r)$, note that we only have contributions from $|z| \ll 1$, so we can write $p_z(z) \simeq (2\pi)^{-1}$. Transforming to spherical polar coordinates $\lambda_0 = u \cos \theta$ and $z\Omega(r) = u \sin \theta e^{i \varphi}$ we find $\omega(0;\lambda_0,z,r) = u$. The result is
\begin{align}
	p_{\omega|r}(\omega',r) \simeq \pi^{-1} [\omega'/\Omega(r)]^2,
\label{eq:pomegarappendix2}
\end{align}
where the quadratic dependence on the splitting comes from the integration measure. In between the regimes $\omega \ll \Omega(r)$ and $\omega \gg \Omega(r)$ it can be verified that $p_{\omega|r}(\omega,r)$ interpolates smoothly between the results in Eqs.~\eqref{eq:pomegarappendix1} and \eqref{eq:pomegarappendix2}.

In Sec.~\ref{sec:twopoint} we have determined $p_{\omega}(\omega)$ from the above distributions of $\omega_{br}$. There we neglected the case $r=1$, which we now discuss briefly. In the decoupled system ($J=0$) each site evolves under an independent Haar random $2 \times 2$ unitary matrix. The distribution of single-site level separations is then $p_{\omega}^{\text{Haar}}(\omega) = (2/\pi)\sin^2(\omega/2)$, which follows from standard properties of Haar measure. In the many-body problem with $J \neq 0$, we expect that pairs of LIOM configurations that differ only on a single site are separated in quasienergy by $\omega$ distributed according to $p_{\omega|r}(\omega,1) \simeq p_{\omega}^{\text{Haar}}(\omega)$. With $L$ sites, we therefore find
\begin{align}
	p_{\omega}(\omega) &= \sum_{r=1}^L p_r(r) p_{\omega|r}(\omega,r) \\
	&\simeq L 2^{-L} p_{\omega}^{\text{Haar}}(\omega) + \sum_{r=2}^L p_r(r) p_{\omega|r}(\omega,r). \notag
\end{align}
In the main text we have neglected the first of the above terms, and this is because $p_{\omega}^{\text{Haar}}(\omega)$ is small for small $\omega$. For $1 < r \leq L/2$ we have $p_r(r)$ from Eq.~\eqref{eq:pr}. Furthermore, for large $L$, $p_r(r)$ is given approximately by Eq.~\eqref{eq:pr} even for $L/2 < r \lesssim L$. If we restrict ourselves to $\Omega(L) \ll \omega \ll 1$, we find Eq.~\eqref{eq:pomega2}.

\section{Distribution of curvatures associated with individual local resonances}\label{app:curvaturedist}
In Sec.~\ref{sec:curvatures} we have calculated the distribution of level curvatures $\kappa$. To do so we required the distribution $p_{\tilde \kappa}(\tilde \kappa)$. In this appendix we discuss $p_{\tilde \kappa}(\tilde \kappa)$ in more detail.

The distribution of $\tilde \kappa$ conditioned on $r$ is given by
\begin{align}
	p_{\tilde \kappa|r}(\tilde \kappa',r) = \int d^2 z p_z(z) p_{\lambda_0}(\lambda_0) \delta\big[\tilde  \kappa' - \tilde \kappa(\lambda;\lambda_0,z,r)\big],
\notag
\end{align}
with $\kappa(\lambda;\lambda_0,z,r)$ given in Eq.~\eqref{eq:tildekappapairwise}; here, as in Eq.~\eqref{eq:kappasum2}, we discard the subscript $br$. We also set $\lambda=0$, as in Appendix~\ref{app:omegadist}. Far from the resonance, with $|\lambda_0| \gg |z|\Omega(r)$, we have $|\tilde \kappa(\lambda;z,r)| \simeq (1/2)|z|^2 \Omega^2(r)/\lambda_0^3$. Within this approximation we obtain the distribution $p_{\tilde \kappa|r}(\tilde \kappa,r)$ for $|\tilde \kappa| \ll 1/\Omega(r)$. Changing integration variable from $\lambda_0$ to $(1/2)|z|^2 \Omega^2(r)/\lambda_0^3$, we find, for $4|z|^2 \Omega^2(r)/\Lambda^3 < |\tilde \kappa| \ll 4/(|z|\Omega)$,
\begin{align}
	p_{\tilde \kappa|r}(\tilde \kappa,r) = \eta \Omega^{2/3}(r) [2\Lambda]^{-1} |\tilde \kappa|^{-4/3},
\end{align}
where the numerical factor $\eta$ is of order unity.

Close to resonances we anticipate $|\tilde \kappa| \sim \Omega^{-1}(r)$. We now show that the probability for $|\tilde \kappa| \gg \Omega^{-1}(r)$ decays rapidly with increasing $|\tilde \kappa|$. So that $|\tilde \kappa| \gg \Omega^{-1}(r)$, we must have $|z| \ll 1$ and $\lambda \ll 1$. In this regime we can therefore approximate $p_z(z)$ by a constant. Switching to spherical polars $z\Omega = u \sin \theta e^{i \varphi}$ and $\lambda_0 = u \cos\theta$ we then find
\begin{align}
	p_{\tilde \kappa|r}(\tilde \kappa',r) = \frac{1}{\Lambda \Omega^2} \int d\theta du u^2 \sin \theta \delta\Big[\tilde  \kappa' - \frac{\sin^2 \theta}{2 u}\Big],
\end{align} 
where we have evaluated the integral over the $\varphi$ coordinate. This leads to
\begin{align}
	p_{\tilde \kappa|r}(\tilde \kappa,r) = \eta' \Lambda^{-1} \Omega^{-2}(r) |\tilde \kappa|^{-4},
\end{align}
for $\eta'$ of order unity.

For $|\tilde \kappa| \ll \Omega^{-1}(r)$ the distribution $p_{\tilde \kappa|r}(\tilde \kappa,r)$ decays as $|\tilde \kappa|^{-4/3}$, whereas for $|\tilde \kappa| \gg \Omega^{-1}(r)$ it decays as $|\tilde \kappa|^{-4}$. There is a crossover between these power laws around $|\tilde \kappa| \sim \Omega^{-1}(r)$. Considering the moments of $|\tilde \kappa|$ for each $r$, we see that $\langle |\kappa|^n \rangle \sim \Omega^{-n}(r)$ for any $1/3 < n < 3$.

To determine the full distribution $p_{\tilde \kappa}(\tilde \kappa)$ we must sum over all possible values of $r$ as in Eq.~\eqref{eq:ptildekappa1}. Contributions from $r=1$ terms arise even for the decoupled system ($J=0$), and make significant contributions only for the smallest $\tilde \kappa$, as suggested by Fig.~\ref{fig:curvature1}. We approximate $p_r(r) = L 2^{r-L}$ [Eq.~\eqref{eq:pr}] for all $r < L$ (this is the exact result only for $r \leq L/2$). For $r \ll \zeta \ln |\tilde \kappa|$, the contributions from $r$-resonances scale as $2^r \Omega^{-2}(r) |\tilde \kappa|^{-4}$. In the opposite extreme of $r \gg \zeta \ln |\tilde \kappa|$, the contributions from $r$-resonances scale as $2^r \Omega^{2/3}(r)|\tilde \kappa|^{-4/3}$. 

Due to the slow decay of $p_{\tilde \kappa|r}(\tilde \kappa,r)$ for $|\tilde \kappa| \ll \Omega^{-1}(r)$, for $\zeta < (2/3)\zeta_c$ we anticipate that $p_{\tilde \kappa}(\tilde \kappa)$ is dominated by $r \sim \zeta \ln |\tilde \kappa|$. For smaller $r$ the probability density at $|\tilde \kappa|$ is suppressed as $|\tilde \kappa|^{-4}$, while for larger $r$ the increase of $p_r(r)$ with $r$ is overwhelmed by the decay of $\Omega^{2/3}(r)$. This implies that
\begin{align}
	p_{\tilde \kappa}(\tilde \kappa) \sim L 2^{-L} |\tilde \kappa|^{-(2-\zeta \ln 2)}.
\end{align}
For $(2/3)\zeta_c < \zeta < 1/\ln 2$ we anticipate different behaviour. In particular, resonances with $r \sim L$ dominate $p_{\tilde \kappa}(\tilde \kappa)$. This is because, for such values of $\zeta$, $2^r \Omega^{2/3}(r)$ is an increasing function of $r$. As a result we find $p_{\tilde \kappa}(\tilde \kappa) \sim |\tilde \kappa|^{-4/3}$ with a different $L$-dependence relative to the regime $\zeta < (2/3)\zeta_c$.

Together, our results indicate that on increasing $\zeta$ the power of the decay of $p_{\tilde \kappa}(\tilde \kappa)$ at large $|\tilde \kappa|$ should decrease from $2$ at $\zeta=0$ to $4/3$ for $(2/3) \zeta_c < \zeta < \zeta_c$. This trend is in good agreement with the numerical results in Fig.~\ref{fig:curvature1}. Our theoretical arguments above also suggest a subtle change in the scaling of $p_{\tilde \kappa}(\tilde \kappa)$ with $L$ in these two regimes, although a detailed exploration would require a wider range of system sizes.

\bibliography{mblparametricrefs}
\end{document}